\documentclass[preprint]{elsarticle}
\usepackage{dsfont}
\usepackage{amsmath,amsfonts,amssymb}
\usepackage{graphicx}
\usepackage[font=footnotesize]{caption}
\usepackage{subcaption}
\usepackage{hyperref}

\newcommand{\vr}{\ensuremath{\vec{r}}}

\newcommand{\drm}{\ensuremath{\mathrm{d}}}
\newcommand{\Sp}{\ensuremath{\mathbb{S}^{2}}}
\newcommand{\sigf}{\ensuremath{\sigma_{\text{f}}}}
\newcommand{\sigt}{\ensuremath{\sigma_{\text{t}}}}

\usepackage{alphalph}

\usepackage{epstopdf}
\usepackage{stmaryrd}
\usepackage{float}
\usepackage{color}
\usepackage{cases}
\usepackage{lineno}

\newcommand{\addition}[1]{\textcolor{black}{#1}}

\hypersetup{
	colorlinks,
	citecolor=black,
	filecolor=black,
	linkcolor=blue,
	urlcolor=black
}
\DeclareGraphicsExtensions{.pdf,.png,.jpg,-eps-converted-to.pdf,.bmp}

\begin{document}

\begin{frontmatter}
\title{Implicit Filtered P$_{N}$ for High-Energy Density Thermal Radiation Transport using Discontinuous Galerkin Finite Elements \tnoteref{nsf-support}}
\date{\today}
\author[tamu-ne]{Vincent M. Laboure\corref{cor}}
\ead{vincent.laboure@tamu.edu}

\author[tamu-ne]{Ryan G.\,McClarren}
\ead{rgm@tamu.edu}

\author[ornl,utk-math]{Cory D.\,Hauck}
\ead{hauckc@ornl.gov}

\cortext[cor]{Corresponding author. Tel.:+1 979 224 8506}

\tnotetext[nsf-support]{This material is based, in part, upon work supported by the National Science Foundation under Grant No. 1217170.  The research of the third author is sponsored by the Office of Advanced Scientific Computing Research; U.S. Department of Energy. The work was performed at the Oak Ridge National Laboratory, which is managed by UT-Battelle, LLC under Contract No. De-AC05-00OR22725. }

\address[ornl]{Computational and Applied Mathematics Group,
	Oak Ridge National Laboratory,
	Oak Ridge, TN 37831 USA }

\address[utk-math]{Department of Mathematics,
	University of Tennessee
	Knoxville, TN 37996-1320}

\address[tamu-ne]{Nuclear Engineering Department,
	Texas A\&M University
	College Station, TX 77843}

\begin{abstract}
	In this work, we provide a fully-implicit implementation of the time-dependent, filtered spherical harmonics (FP$_{N}$) equations for non-linear, thermal radiative transfer.  
	We investigate local filtering strategies and analyze the effect of the filter on the conditioning of the system, showing in particular that the filter improves the convergence properties of the iterative solver.  
	We also investigate numerically the rigorous error estimates derived in the linear setting, to determine whether they hold also for the non-linear case. 
	Finally, we simulate a standard test problem on an unstructured mesh and make comparisons with implicit Monte-Carlo (IMC) calculations.
\end{abstract}

	
\end{frontmatter}



\section{Introduction}

The equations of thermal radiative transfer describe the movement of photons through a material as well as the exchange of energy between the photon radiation and the material.  
There are two equations:  a radiation transport equation that tracks the energy in the radiation field via an angular intensity $\mathcal{I}$ and a temperature equation that tracks the internal energy of the material.  
The coupling of these two equations reflects the exchange of energy as photons are emitted and absorbed by the material.

Various numerical methods for the radiation transport equation have been developed to solve the radiative transfer problem.  
The challenge here is that the angular intensity is, in the most general setting, a function of six phase space variables (position, energy, and direction of propagation) plus time.  
The most common approaches are
implicit Monte Carlo methods \cite{FleckCummings1971,McClarren2009}, discrete ordinate methods \cite{LarsenMorel2010}, spectral approximations \cite{Brunner2005}, finite element discretizations \cite{Kanschat-2009}, and nonlinear moments methods \cite{EntropyBasedClosures,Hau10,Dubroca-Feugas-1999}.   

In this paper, we focus on a variation of the spherical harmonics (or  P$_{N}$) method.  The P$_{N}$ method is a spectral Galerkin method that approximates the angular dependence of the radiation intensity using a finite expansion in spherical harmonics up to degree $N$.  
The result is a linear, hyperbolic system of time-dependent equations for the expansion coefficients, which can then be discretized with respect to space and time in a variety of ways.

The P$_{N}$ approach offers several benefits.  Among these are spectral convergence for smooth solutions and preservation of the rotational invariance of the transport equation.%
\footnote{Roughly speaking, the solution of the equation is unchanged when the spatial and angular variables in phase space undergo the same rotation.}  
However, the method also poses challenges.  
Chief among these is that the P$_{N}$ approximation of the angular intensity can be highly-oscillatory and even negative when the underlying exact solution is not sufficiently smooth; this happens typically in regions where the material cross-section is small. In addition to being non-physical, the negative radiation energy can cause the material temperature $T$ to become negative, in which case the model for photon emission is not well-defined.%
\footnote{For $T \leq 0$, the expression of the Planckian -- as shown in Eq.\,\ref{planckianEq} below -- is not integrable with respect to energy.}  
In addition, the material cross-section may become negative and thereby introduce instabilities into the simulation.

To address the problem of Gibbs phenomena in the P$_{N}$ approximation, McClarren and Hauck \cite{McClarrenHauck2010} applied filtering techniques to smooth out the angular dependency of the solution; they coined the name \textit{filtered P$_{N}$} (FP$_{N}$) for this method. 
While filtering does not ensure positivity of the approximation for $\mathcal{I}$, it does suppress oscillations in the P$_{N}$ approximation at a computational cost that is much lower than other closures that are robustly positive, such as positive P$_{N}$ (PP$_N$) closures \cite{Garrett-Hauck-2015,HauckMcClarren2010,Laiu-Hauck-McClarren-Oleary-Tits-2016} and entropy-based closures \cite{Hau10,Dubroca-Feugas-1999}. 
In practice, the filtering approach has so far shown very promising results \cite{McClarrenHauck2010,PhysicsLettersA2010,mcclarren2008filtered,ahrens2013,Radice2013}.
However, the method has yet to be implemented with an implicit time-integration scheme, which is often preferred due to the extremely fast scales in the transport equation.  
Indeed, in an explicit scheme, the time step $\Delta t$ required by particle advection is bounded by $\Delta x/c$, where $\Delta x$ characterizes the size of the spatial mesh and $c$ is the speed of light.%
\footnote{An explicit treatment of the energy exchange terms may require an even smaller time step.
	However, because these terms are spatially local, they are relatively easy to treat implicitly.}
Such a condition is often too restrictive.  
Implicit methods, on the other hand, maintain stability with a much larger time step. However, each step requires the inversion of a large set of algebraic equations. 

In this paper, we detail an implicit implementation of the filtered P$_{N}$ equations using Discontinuous Galerkin (DG) Finite Elements.  
The DG approach is one of several possible spatial discretization methods.  
Other finite element approaches for P$_{N}$ include least-squares formulations \cite{GeshDissertation, manteuffel1998least, manteuffel1999boundary}, parity-based formulations \cite{HerbertEgger2012, Wright-Arridge-Schweiger}, self-adjoint formulations \cite{MorelMcGhee1999}, and streamlined-upwind Petrov-Galerkin methods \cite{pain2006streamline,pain2006space}.
Discontinuous Galerkin methods were invented for transport problems in Ref.\,\cite{reed1973triangularmesh}.
There it was observed that the discontinuous basis, while more expensive than a standard continuous approximation, give better approximations to problems with non-smooth solutions.
In addition to being robust in streaming regimes, where non-smooth solutions typically occur, DG methods (with a sufficiently rich basis set) also perform well in the diffusion limit \cite{GuermondKanschat2010, LarsenMorelMiller2001, AdamsDFE}.%
\footnote{Roughly speaking, this limit occurs when particle interactions with the surrounding medium isotropize the radiation field and the angular average of the photon distribution satisfies a much simpler diffusion equation \cite{Habetler-Matkowsky-1975,Larsen-Keller-1974}.}
A semi-implicit discretization of the P$_N$ equations with DG methods, which treats the flux terms explicitly, can be found in Ref.\,\cite{McClarrenEvansLowrie2008}.

The remainder of this paper is organized as follows. 
The radiative transfer equations and the FP$_N$ equations are presented in Section \ref{Sec2}.  
The spatial discretization of the FP$_N$ equations is presented in Section \ref{Sec3}.
In Section \ref{Sec4}, we show the impact the filter has on the convergence properties of the iterative solver and then consider the error estimates derived in Ref.\,\cite{FranckHauck2014} for the linear setting.
Finally, in Section \ref{Sec5}, we test the method with different filtering strategies on the challenging benchmark problem known as the Crooked Pipe \cite{CrookedPipe}.  Because this problem is particularly hard to converge, we first show good agreement between our code and implicit Monte-Carlo calculations on a simplified version.  We then show for the harder problem that the filter mitigates deficiencies in the P$_N$ solutions, especially for smaller values of $N$.

\section{Implicit Filtered P$_N$}
\label{Sec2}
We consider the grey (frequency integrated) form of the thermal radiative transfer equations, given by
\cite{pomraning1973equations}:
\begin{equation}
\dfrac{1}{c} \dfrac{\partial \mathcal{I}}{\partial t} +\vec{\Omega}\cdot \vec{\nabla} \mathcal{I} +\sigma_{\mathrm{t}}(T) \mathcal{I} = \sigma_{\mathrm{a}}(T) B(T) + \dfrac{\sigma_{\mathrm{s}}}{4\pi} \phi +\mathcal{Q},
\label{eq0}
\end{equation}
\begin{equation}
\dfrac{\partial}{\partial t} E(T)  = \sigma_{\mathrm{a}}(T)\big(\phi - 4\pi B(T)\big),
\label{eq01}
\end{equation} 
along with appropriate initial and boundary conditions. Eq.\,\ref{eq0} governs the angular intensity  $\mathcal{I}(\vec{r},\vec{\Omega},t)$ of the photon radiation, with $\vec{r}$ and $\vec{\Omega}$ being, respectively, the spatial and angular coordinates and $t$ being the time.  Meanwhile, Eq.\,\ref{eq01} governs the evolution of the material energy $E(T)$, where $T(\vec{r}, t)$ is the material temperature. The derivative $C_{v} = E'(T)$ is the material heat capacity; for calculations, we assume it is independent of $T$, although the formulations do not require it.  The constant $c$ is the speed of light; $\sigma_{\mathrm{s}}$, $\sigma_{\mathrm{a}}$, and $\sigma_{\mathrm{t}}= \sigma_{\mathrm{s}}+\sigma_{\mathrm{a}}$ are the scattering, absorption, and total macroscopic cross-sections, respectively, with units of inverse length; $\mathcal{Q}$ is the (known) volumetric source. The scalar intensity $\phi(\vec{r},t)$ is the integral of the specific intensity with respect to angle  
\begin{equation}
\phi \equiv \int_{\mathbb{S}^{2}} \mathcal{I}\,\mathrm{d}\Omega,
\end{equation}
where $\mathbb{S}^{2}$ is the unit sphere; the frequency-integrated Planckian blackbody source is given by
\begin{equation}
B(T) \equiv\int_{0}^{\infty} \,\dfrac{2h\nu^{3}}{c^{2}}\dfrac{1}{\exp(\frac{h\nu}{kT})-1} \mathrm{d}\nu = \dfrac{acT^{4}}{4\pi},
\label{planckianEq}
\end{equation}
where $a = {(8\pi^5k^4)}/{(15h^3c^3)}$ is the radiation constant, with $h$ and $k$ being the Planck and Boltzmann constants, respectively. This integral is only defined for $T>0$, which is one reason to maintain a positive material temperature.

We assume that Eq.\,\ref{eq0} is defined over a bounded spatial domain $\mathcal{D}$ and we let $\mathcal{S} = \mathcal{D} \times \mathbb{S}^2$.  Boundary conditions for $\mathcal{I}$ must be specified for incoming data---that is, on the set
\begin{equation}
\partial \mathcal{S} ^- = \{(\vec{r},\vec{\Omega}) \in \partial \mathcal{D}\times \Sp : \vec{n}_0(\vec{r}) \cdot \vec{\Omega} < 0 \},
\end{equation}
where $\vec{n}_0(\vec{r})$ is the outward normal at a point $\vec{r} \in \partial \mathcal{D}$.

\subsection{Fully-implicit radiation transfer}
Applying the backward Euler method to discretize Eqs.\,\ref{eq0} and \ref{eq01} in time leads to the following quasi-steady form of the radiative transfer system:
\begin{equation}
\vec{\Omega}\cdot \vec{\nabla} \mathcal{I}^{n+1} +\sigma_{\mathrm{t}}^{*}(T^{n+1}) \mathcal{I}^{n+1} = \sigma_{\mathrm{a}}(T^{n+1}) B(T^{n+1}) + \dfrac{\sigma_{\mathrm{s}}}{4\pi} \phi^{n+1} +\mathcal{Q}^{*},
\label{eq2}
\end{equation}
\begin{equation}
\dfrac{E(T^{n+1}) - E(T^{n}) }{\Delta t} = \sigma_{\mathrm{a}}(T^{n+1})\Big(\phi^{n+1} - 4\pi B(T^{n+1})\Big),
\label{eq3}
\end{equation}
where
\begin{equation}
\sigma_{\mathrm{t}}^{*} =  \sigma_{\mathrm{t}} + \dfrac{1}{c\Delta t}\quad \text{and}\quad
\mathcal{Q}^{*} =  \frac{1}{\Delta t}\int_{t^n}^{t^{n+1}}\mathcal{Q}\,\drm t + \dfrac{\mathcal{I}^{n}}{c\Delta t}.
\label{eqQstar}
\end{equation}
Here and throughout, the superscript $n$ indicates the discrete approximation of a time-dependent quantity at time $t^n$. When a superscript is not specified, it is assumed that such approximations are evaluated at $t^{n+1}$.

The system \ref{eq2}-\ref{eq3} is nonlinear due to the Planckian term $B$ and possibly the material properties. In this work, we choose a fully nonlinear treatment although it has been shown that expanding $B$ about $T^n$ and evaluating the cross-sections at the previous time step also performs well \cite{Lowrie2004,McClarrenEvansLowrie2008}.

\subsection{Spherical Harmonics Expansion of the transport equation}
In the P$_N$ equations, $\mathcal{I}$ is approximated by a finite spherical harmonic expansion:%
\footnote{Even though $\hat{\mathcal{I}}$ depends on both $n$ and $N$, we omit these dependencies in order to simplify the notation.}
\begin{equation}\label{eq:Iapprox}
\mathcal{I}^{n+1}(\vec{r},\vec{\Omega}) \approx \hat{\mathcal{I}}(\vec{r},\vec{\Omega}) = \sum_{\ell=0}^{N}\sum_{m=-\ell}^{\ell} I_{\ell}^{m}(\vec{r})\,R_{\ell}^{m}(\vec{\Omega}),
\end{equation}
where, for variables $\mu \in [-1,1]$ and $\phi \in [0, 2 \pi)$ such that $\vec{\Omega} = \sqrt{1-\mu^{2}}\cos\varphi\,\vec{e}_x+\sqrt{1-\mu^{2}}\sin\varphi\,\,\vec{e}_y +\mu\,\vec{e}_z$, the real-form spherical harmonics are given by:
\begin{equation}
R_{\ell}^{m}(\vec{\Omega}) = 
\begin{cases}
\sqrt{2}\,C_{\ell}^{m}\,P_{\ell}^{m}(\mu)\cos(m\varphi), & 0 < m \leq \ell \leq N
\\
C_{\ell}^{0}\,P_{\ell}^{0}(\mu), & 0 \leq \ell \leq N
\\
\sqrt{2}\,C_{\ell}^{|m|}\,P_{\ell}^{|m|}(\mu)\sin(|m|\varphi), & 0 < -m \leq \ell \leq N
\end{cases}  .
\end{equation}  
Here $C_{\ell}^{m} = \sqrt{\frac{(2\ell+1)}{4\pi}\frac{(\ell-m)!}{(\ell+m)!}}$ is a normalization constant chosen such that $\int_{\mathbb{S}^{2}} R_{\ell}^{m} R_{\ell'}^{m'}\drm\Omega = \delta_{\ell,\ell '}\delta_{m, m'}$, with $\delta_{\ell,\ell'}$ being the Kronecker delta, and $P_{\ell}^{m}$ denotes the associated Legendre polynomial of degree $\ell$ and order $m$.%

Integrating Eq.\,\ref{eq2} in angle against $R_{\ell}^{m}$ and applying the approximation in Eq.\,\ref{eq:Iapprox} gives, for all $(\ell,m)\in\mathcal{N} \equiv \{(\ell,m)\in\mathbb{N}^{2}:0\leq|m|\leq \ell\leq N\}$,
\begin{equation}
\int_{\mathbb{S}^{2}} \vec{\Omega}\cdot \vec{\nabla} \hat{\mathcal{I}} \;R_{\ell}^{m}\,\mathrm{d}\Omega + \sigma_{\mathrm{t}}^{*}\,  I_{\ell}^{m} - \sigma_{\mathrm{s}} I_{0}^{0}\delta_{\ell,0} = \sqrt{4\pi}\sigma_{\mathrm{a}}B \delta_{\ell,0}  +Q_{\ell}^{m}, 
\label{eq16}
\end{equation}
where $
Q_{\ell}^{m} = \int_{\mathbb{S}^{2}}\,\mathcal{Q}^{*}\,R_{\ell}^{m}\,\mathrm{d}\Omega.
$
In Eq.\,\ref{eq16}, the angular moments are coupled to each other only through the streaming operator $\vec{\Omega}\cdot \vec{\nabla}$. This coupling is expressed through the matrices
\begin{equation}
A_{\chi} \equiv \int_{\mathbb{S}^{2}} \vec{\Omega}\cdot \vec{e}_{\chi}\, \mathcal{R}\,\mathcal{R}^T \,\mathrm{d}\Omega\quad,\quad \chi\in\{x,y,z\},
\end{equation}
where $\mathcal{R}$ is the vector containing the spherical harmonics $R_{l}^{m}$. These matrices can be evaluated using well-known recursion relations (see \cite{FranckHauck2014,Radice2013} or \cite{Brunner2000} for the complex version) or exact quadrature rules.

We collect the expansion coefficients $I_{\ell}^m$ into a vector $I$ using a consistent ordering with a single index, and write Eq.\,\ref{eq16} as the following linear system: 
\begin{equation}
A_{x}\dfrac{\partial I}{\partial x} + A_{y}\dfrac{\partial I}{\partial y} + A_{z}\dfrac{\partial I}{\partial z}+ \sigma_{\mathrm{t}}^{*}  I - \sigma_{\mathrm{s}} \varPhi
= \big(\sqrt{4\pi}\sigma_{\mathrm{a}}B\big)\mathds{1}  +Q,
\label{eq1}
\end{equation}
where $\mathds{1} = (1 \quad 0 \quad \cdots \quad 0)^T$ and $\varPhi =(I_{0}^{0} \quad 0 \quad\cdots\quad 0)^T$.  In a slight abuse of notation, we denote the $j^{\rm{th}}$ component of $I$ in the single index ordering by $I_j = I_{\ell}^m$, where the map $(\ell,m) \leftrightarrow j$ is a bijection between the two sets of indices.%
\footnote{We have thus assumed that $(\ell,m) = $ (0,0) is associated to $j=$ 1.}
The same convention will be used for $Q_{\ell}^{m}$.

The solution vector $I$ has $(N+1)^{2}$ components, but in reduced geometries, there are only $ P < (N+1)^{2}$ that are not redundant or trivially zero. If $\mathcal{I}$ depends on only two spatial dimensions, then $P = \frac12(N+1)(N+2)$; if $\mathcal{I}$ depends on only one spatial dimension, then $P=N+1$ \cite{Brunner2000}.%
\footnote{In practice, we solve only for the nontrivial moments, but for simplicity we maintain the notations in Eq.\,\ref{eq1} even when $P< (N+1)^{2}$.} 

\subsection{Angular filtering}
\label{sec2}
The purpose of angular filtering is to reduce unphysical oscillations that can arise from truncating the spherical harmonics expansion.  It has been demonstrated theoretically in \cite{McClarren2008} and observed numerically in \cite{HauckMcClarren2010,Brunner2002, GarrettHauck2013} that these oscillations may lead to negative solution values for the scalar intensity $\phi$.  In its original implementation, the filter suppresses them by damping higher-order angular moments $(\ell >0)$ after each time step in the given temporal integration scheme and, in doing so, effectively mitigates negative scalar intensities in the P$_N$ solution. 
In addition, the filter was constructed in such a way as to conserve energy, preserve rotational invariance, and maintain formal convergence of the solution as $N$ goes to infinity.

Radice et al.\cite{Radice2013} later showed that with an appropriate modification of the filter strength, one can derive a modified set of equations. 
In effect, their formulation adds artificial scattering to the system, replacing Eq.\,\ref{eq1} by:
\begin{equation}
A_{x}\dfrac{\partial I}{\partial x} + A_{y}\dfrac{\partial I}{\partial y} + A_{z}\dfrac{\partial I}{\partial z}+ \sigma_{\mathrm{t}}^{*}  I +\sigf D I -\sigma_{\mathrm{s}} \varPhi
= \big(\sqrt{4\pi}\sigma_{\mathrm{a}}B \big)\mathds{1}  +Q,
\label{eq11}
\end{equation}
where $\sigf$ is a free parameter, $(D I)_{\ell}^{m} = f(\ell,N)\,I_{\ell}^{m}$, and
the \textit{filter function} $f$ is given by:
\begin{equation}
\label{eqFilter}
f(\ell,N) \equiv  - \log \rho_{\text{filterType}}\left(\dfrac{\ell}{N+1}\right) .
\end{equation}
The Lanczos and spherical spline filters are considered \cite{Radice2013}:
\begin{equation}
\rho_{\text{Lanczos}}(\zeta) = \dfrac{\sin\zeta}{\zeta}\\
\,;\quad
\rho_{\text{SSpline}}(\zeta) = \dfrac{1}{1+\zeta^{4}}
.
\label{filterDef}
\end{equation}
%

The variable $\sigf$ in Eq.\,\ref{eq11} is a tuning parameter -- henceforth called \textit{filter strength} -- that may be spatially dependent. Strategies for determining a good local value of $\sigf$ are discussed in Section \ref{SecFilteringStrategy}. In this context, one of the strengths of the reformulation in \cite{Radice2013} is that --- unlike the original implementation in \cite{McClarrenHauck2010} --- the filter strength is independent of the size of the time step and the spatial mesh \cite{Radice2013}.  Thus the value of $\sigf$ needs to be tuned only once, and this can be done using relatively cheap simulations on coarse meshes.

\section{Spatial Discretization}
\label{Sec3}

We discretize Eq.\,\ref{eq11} in space, along with Eq.\,\ref{eq3} for the material temperature, using the Discontinuous Galerkin Finite Element Method (DGFEM).  This method is, by now, fairly standard.  Thus the presentation here will be brief.  Roughly speaking, the method relies on a piecewise polynomial approximation of the true solution.  It requires the specification of a numerical flux at points of discontinuities, the effect of which is to add a stabilizing term to the usual variational form.  For details, we refer the reader to the review in \cite{DG2001}.

\subsection{Variational formulation}
\label{varFormulation}

Let $\mathcal{T}$ be a collection of open convex, polyhedral cells $K \subset \mathcal{D}$ such that $\cup \overline{K} = \mathcal{D}$, and let $h>0$ be the size of the largest disk that can be inscribed inside any cell $K$. Let $\Gamma_{\text{int}}$ be the set of interior facets:
\begin{equation}
\Gamma_{\text{int}} = \{ e: e = \overline{K}_1 \cap \overline{K}_2 ~\text{for any}~ K_1,K_2 \in \mathcal{T},~K_1\neq K_2 \} \:.
\end{equation}
Let $V$ be a finite-dimensional trial space of functions that are polynomial on each $K \in \mathcal{T}$. For each $j$, we seek a function $I^h_j \in V$ that approximates the coefficient $I_j = I_l^m$.  
Thus the approximation $I^h$ of the vector-valued function $I$ lives in the (Cartesian) product space $V^P$. We also approximate $T$ by $T^h \in V$.  

The formulation of the DGFEM is as follows:  Find $(I^h,T^h) \in V^{P}\times V$ such that  $a((I^h,T^h),v) = L(v)$ for all $v\in V^{P+1}$, where for each $((u,\theta),v) \in (V^{P}\times V) \times V^{P+1}$,

\begin{equation}
a((u,\theta),v) = \sum_{i=1}^{P+1} a_{i}((u,\theta),v_{i}) \qquad \text{and} \qquad
L(v) = \sum_{i=1}^{P+1} L_{i}(v_{i}).
\end{equation}
Here $a_i$ and $L_i$ denote (for each $i$, $1\leq i \leq P$) forms associated to the $i$-th equation of Eq.\,\ref{eq11}, and to Eq.\,\ref{eq3} for $i=P+1$. 
They are derived by multiplying the corresponding equation by $v_i$ and integrating over each cell $K \in \mathcal{T}$. For $1\leq i\leq P$, following an integration by parts, $a_i$ can expressed as the sum of four terms:
\begin{equation}
a_i((u,\theta),v_i) = a_i^{\text{vol}}((u,\theta),v_i) + a_i^{\text{int}}(u,v_i) + a_i^{\text{ext}}(u,v_i) + a_i^{\text{BC}}(u,v_i).
\label{varForm}
\end{equation}
Here $a_i^{\text{vol}}$ is the volumetric contribution; $a_i^{\text{int}}$ is the contribution from interior facets; $a_i^{\text{ext}}$ is a contribution from exterior facets (along the boundary of the spatial domain $\partial\mathcal{D}$); and $a_i^{\text{BC}}$ is a boundary contribution.

\begin{figure}[h]
	\centering
	\includegraphics[scale=0.3]{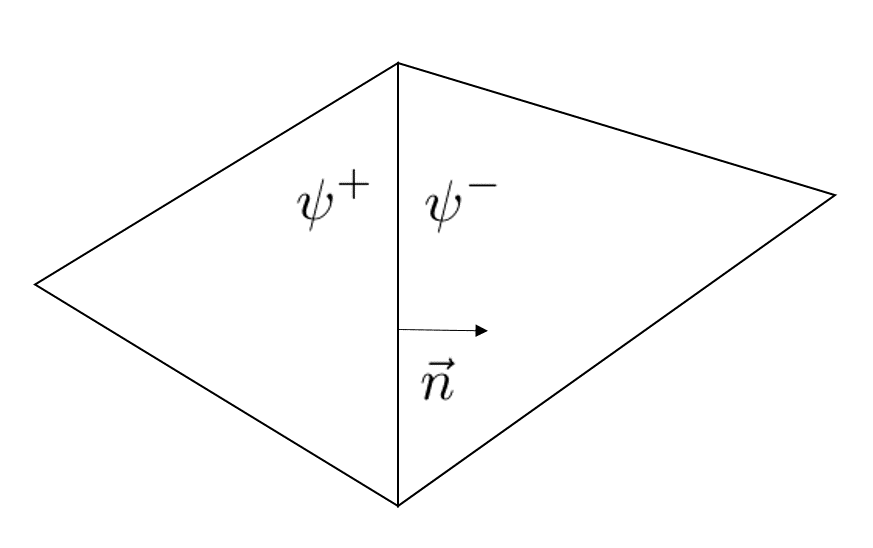}
	\caption{Notation for discontinuous variables, given a unit normal vector $\vec{n}$.}
	\label{fig1}
\end{figure}

The term $a_i^{\text{BC}}$ accounts for any boundary conditions that express incoming information in terms of outgoing information (such as reflective boundaries) and will be discussed in Section \ref{BCsec}. The remaining terms are\footnote{Recall that we have assumed that $v_1$ is the test function associated to the 0-th moment equation of Eq.\,\ref{eq11} and that $v_{P+1}$ is the test function associated to the temperature equation (Eq.\,\ref{eq3}).}
\begin{equation}
\begin{split}
a_i^{\text{vol}}((u,\theta),v_i) = &  \bigg(\sum_{j = 1}^{P} \int_{ \mathcal{D}}\,\big((\sigma_{\mathrm{t}}^{*}\delta_{ij} + \sigf D_{ij})v_{i} - A_{x,ij}\dfrac{\partial v_{i}}{\partial x} - A_{y,ij}\dfrac{\partial v_{i}}{\partial y} - A_{z,ij}\dfrac{\partial v_{i}}{\partial z} \big) u_{j} \, \mathrm{dx} \bigg)\\
& - \int_{ \mathcal{D}}\, \left(\sigma_{\mathrm{s}}\, u_{1}  + \sqrt{4\pi}\sigma_{\mathrm{a}}B\right) v_{1}\,\delta_{i,1}\, \mathrm{dx}
,
\end{split}
\end{equation}
\begin{equation}
\begin{split}
a_i^{\text{int}}(u,v_i) = \sum_{j = 1}^{P}\sum_{e\in\Gamma_{\text{int}}} \bigg(&\int_{e}\,\Big(\vec{e}_{x}\cdot \vec{n}\;A_{x,ij}\, \llbracket v_{i} \rrbracket \langle u_{j}\rangle +\dfrac{1}{2}|\vec{e}_{x}\cdot \vec{n}\,|\,M_{x,ij}\, \llbracket v_{i} \rrbracket \llbracket u_{j} \rrbracket\Big) \mathrm{ds} \\
+&\int_{e}\,\Big(\vec{e}_{y}\cdot \vec{n}\;A_{y,ij}\, \llbracket v_{i} \rrbracket \langle u_{j}\rangle +\dfrac{1}{2}|\vec{e}_{y}\cdot \vec{n}\,|\,M_{y,ij}\, \llbracket v_{i} \rrbracket \llbracket u_{j} \rrbracket\Big) \mathrm{ds} \\
+&\int_{e}\,\Big(\vec{e}_{z}\cdot \vec{n}\;A_{z,ij}\, \llbracket v_{i} \rrbracket \langle u_{j}\rangle +\dfrac{1}{2}|\vec{e}_{z}\cdot \vec{n}\,|\,M_{z,ij}\, \llbracket v_{i} \rrbracket \llbracket u_{j} \rrbracket\Big) \mathrm{ds}
\bigg),
\end{split}
\label{eq_a_vol}
\end{equation}
\begin{equation}
\begin{split}
a_i^{\text{ext}}(u,v_i) = \sum_{j = 1}^{P} \bigg(&\int_{\partial \mathcal{D}}\,\dfrac{1}{2}\Big(\vec{e}_{x}\cdot \vec{n}_{0}\;A_{x,ij} + |\vec{e}_{x}\cdot \vec{n}_{0}\,|\,M_{x,ij}\Big)v_{i}\, u_{j}\, \mathrm{ds} \\
+&\int_{\partial \mathcal{D}}\,\dfrac{1}{2}\Big(\vec{e}_{y}\cdot \vec{n}_{0}\;A_{y,ij} + |\vec{e}_{y}\cdot \vec{n}_{0}\,|\,M_{y,ij}\Big)v_{i}\, u_{j}\, \mathrm{ds} \\
+&\int_{\partial \mathcal{D}}\,\dfrac{1}{2}\Big(\vec{e}_{z}\cdot \vec{n}_{0}\;A_{z,ij} + |\vec{e}_{z}\cdot \vec{n}_{0}\,|\,M_{z,ij}\,\Big)v_{i}\, u_{j}\, \mathrm{ds}
\bigg),
\end{split}
\label{a_ext}
\end{equation}
Here $\vec{n}$ is a unit vector normal to the interior facet; $\vec{n}_{0}$ is the outward\footnote{The outward direction is defined with respect to $\mathcal{D}$.} unit normal vector on the domain boundary; and the exact form of the dissipation matrices $M_x,\,M_y$ and $M_z$ depends on the choice of numerical flux. In this paper, we use a global Lax-Friedrich flux:
\begin{equation}
M_x = M_y = M_z = \lambda \mathds{I},
\end{equation}
with $\lambda = 1$. This form of numerical flux was chosen over the upwind flux, as was used in \cite{Brunner2000}, because it generates significantly fewer non-zero terms in the variational formulation.  The average operator $\langle\cdot\rangle$ and jump operator $\llbracket \cdot \rrbracket$ are defined at any facet for any variable $\psi$ by:
\begin{equation}
\llbracket \psi \rrbracket \equiv (\psi^{+}-\psi^{-}) \quad,\quad \langle\psi\rangle \equiv \dfrac{\psi^{+}+\psi^{-}}{2}\quad,
\end{equation}
with $\psi^{+}$ and $\psi^{-}$ being defined with respect to the unit normal on the facet, cf.\,Fig.\,\ref{fig1}.

\noindent
For each $i$, $1\leq i\leq P$, the linear form $L_i$ is given by
\begin{equation}
\label{linearSource}
L_i(v_i) =   
\int_{ \mathcal{D}}\,Q_{i}\,v_{i} \, \mathrm{dx} 
+ L_i^{\text{BC}}(v_i).
\end{equation}
Here $L_i^{\text{BC}}$ accounts for the boundary conditions and will discussed along with $a_i^{\text{BC}}$ in Section \ref{BCsec}. For $i=P+1$ (i.e.\,the terms associated to Eq.\,\ref{eq3}), we have
\begin{equation}
\label{bilinearTemp}
a_{P+1}((u,\theta),v_{P+1}) 
= \int_{ \mathcal{D}}\, \left(\dfrac{E(\theta)}{\Delta t} 
- \sigma_{\mathrm{a}} \left(\sqrt{4\pi}u_{1} 
- 4\pi B\right) \right) v_{P+1}\, \mathrm{dx},
\end{equation}
\begin{equation}
\label{linearTemp}
L_{P+1}(v_{P+1}) =   \int_{ \mathcal{D}}\, \dfrac{E(T^n)}{\Delta t}\,  v_{P+1}\, \mathrm{dx}.
\end{equation}

\subsection{Mass matrix lumping}
\label{SecLumping}
For robustness in optically thick regions, it may be necessary to lump the matrices corresponding to the collision terms. This was demonstrated in \cite{AdamsDFE} in the context of discontinuous Galerkin discretizations of discrete ordinate equations. In practice, lumping a matrix is done by replacing it by a diagonal matrix whose $i$-th term is the sum of the elements on the $i$-th row of the original matrix. For the Crooked Pipe test problem (see Section \ref{Sec5}) this lumping proved to be necessary to avoid non-physical instabilities in the solution.

\subsection{Initial and boundary conditions}
\label{BCsec}
Initial conditions for $I^h$ and $T^h$ are derived by projecting the initial data for $\mathcal{I}$ and $T$ onto $V^P$ and $V$, respectively.  Boundary conditions are required for $V_h$, but not $T^h$.  Unfortunately, the conditions for $I^h$ cannot be derived directly from the boundary conditions for $\mathcal{I}$, since the former require full moment information and the latter are specified only for incoming data. The boundary conditions that apply to our system are natural, i.e.\,they are imposed weakly in the variational form by adding appropriate terms to the forms $a$ and $L$.
We impose incoming Dirichlet and reflective boundary conditions on $\mathcal{B}_\mathrm{d}$ and $\mathcal{B}_\mathrm{r}$, respectively, where $\mathcal{B}_\mathrm{d} \cup \mathcal{B}_\mathrm{r} = \partial\mathcal{D}$. For $\chi\in\{\mathrm{d},\mathrm{r}\}$, we define $\mathbf{B}^-_{\chi} = \{(\vec{r},\vec{\Omega}) \in \mathcal{B}_{\chi} \times\mathbb{S}^{2}:\vec{n}_0(\vec{r}) \cdot  \vec{\Omega}< 0 \}$. The boundary conditions can then expressed as:
\begin{numcases}{}
\forall (\vec{r},\vec{\Omega}) \in \mathbf{B}^-_\mathrm{d},\,\mathcal{I}(\vec{r},\vec{\Omega}) = g(\vec{r},\vec{\Omega}), \label{Dirichlet} \\
\forall (\vec{r},\vec{\Omega}) \in \mathbf{B}^-_\mathrm{r},\,\mathcal{I}(\vec{r},\vec{\Omega}) = \mathcal I(\vec{r},\vec{\Omega} -2 (\vec{\Omega}\cdot\vec{n}_0)\,\vec{n}_0),
\label{Reflective}
\end{numcases}
where $g$ is given.  Then $a_i^{\text{BC}} = a_i^{\text{BC},\mathrm{d}} + a_i^{\text{BC},\mathrm{r}}$, where each term are described below.

\paragraph{Incoming Dirichlet boundary}  

Dirichlet conditions are imposed by setting values to the incoming data of $I_j$ on the outward side of the exterior facets. The outgoing data is obtained by continuity, that is using the outgoing data of $I_j$ on the inward side of the exterior facets. The numerical flux (still using a Lax-Friedrich flux) can then be defined on $\mathcal{B}_\mathrm{d}$ as:
\begin{equation}
\begin{split}
\mathcal{F}(u,g)  & = \dfrac{\vec{e}_{x}\cdot \vec{n}_0}{2} \left(A_{x}\, u + H_{\text{flux},x}^{\oplus}\,u + g_{\text{flux},x}^{\ominus}\right) + \dfrac{|\vec{e}_{x}\cdot \vec{n}_0|}{2} \left( u -H^{\oplus}\,u - g^{\ominus} \right)   \\
& + \dfrac{\vec{e}_{y}\cdot \vec{n}_0}{2} \left(A_{y}\, u + H_{\text{flux},y}^{\oplus}\,u + g_{\text{flux},y}^{\ominus} \right) + \dfrac{|\vec{e}_{y}\cdot \vec{n}_0|}{2} \left( u -H^{\oplus}\,u - g^{\ominus}\right)  \\
& + \dfrac{\vec{e}_{z}\cdot \vec{n}_0}{2} \left(A_{z}\, u + H_{\text{flux},z}^{\oplus}\,u + g_{\text{flux},z}^{\ominus} \right) + \dfrac{|\vec{e}_{z}\cdot \vec{n}_0|}{2} \left( u -H^{\oplus}\,u - g^{\ominus} \right)  
\end{split}
\label{eqFlux}
\end{equation}
where we have defined the following half-range integrals for all $\chi\in\{x,y,z\}$:
\begin{equation}
g_{\text{flux},\chi}^{\ominus} \equiv \int_{\mathbb{S}^{-}} \vec{\Omega}\cdot \vec{e}_{\chi}\, \mathcal{R}\,g \,\mathrm{d}\Omega\quad,\quad H_{\text{flux},\chi}^{\oplus} \equiv \int_{\mathbb{S}^{+}} \vec{\Omega}\cdot \vec{e}_{\chi}\, \mathcal{R}\,\mathcal{R}^T \,\mathrm{d}\Omega,
\end{equation}
\begin{equation}
g^{\ominus} \equiv \int_{\mathbb{S}^{-}}  \mathcal{R}\,g \,\mathrm{d}\Omega\quad,\quad H^{\oplus} \equiv \int_{\mathbb{S}^{+}} \mathcal{R}\,\mathcal{R}^T \,\mathrm{d}\Omega,
\end{equation}
where $\mathbb{S}^{\pm}(\vec{r}) = \{\vec{\Omega}\in\mathbb{S}^{2}:\pm\vec{n}_0(\vec{r}) \cdot  \vec{\Omega}> 0 \}$ for all $\vec{r}\in\partial\mathcal{D}$.
If $\vec{n}_0(\vec{r})$ is colinear to $\vec{e}_x$, $\vec{e}_y$ or $\vec{e}_z$, the matrices $H_{\text{flux},\chi}^{\oplus}$ and $H^{\oplus}$ can be evaluated exactly using an $(N+1)$-point Gauss-Jacobi quadrature rule. If not, they can be derived using rotation matrices and then applying the quadrature. According to Eq.\,\ref{eqFlux}, the boundary contribution to $a$ (cf.\,Eq.\,\ref{varForm}) is
\begin{equation}
\begin{split}
a_i^{\text{BC},\mathrm{d}}(u,v_i) = \sum_{j = 1}^{P} \bigg(&\int_{ \mathcal{B}_\mathrm{d}}\,\dfrac{1}{2}\Big(H_{\text{flux},x,ij}^{\oplus}\,\vec{e}_{x}\cdot \vec{n}_0 - H_{ij}^{\oplus}\,|\vec{e}_{x}\cdot \vec{n}_0|\Big)v_{i}\, u_{j}\, \mathrm{ds} \\
+&\int_{ \mathcal{B}_\mathrm{d}}\,\dfrac{1}{2}\Big(H_{\text{flux},y,ij}^{\oplus}\,\vec{e}_{y}\cdot \vec{n}_0 - H_{ij}^{\oplus}\,|\vec{e}_{y}\cdot \vec{n}_0|\Big)v_{i}\, u_{j}\, \mathrm{ds} \\
+&\int_{ \mathcal{B}_\mathrm{d}}\,\dfrac{1}{2}\Big(H_{\text{flux},z,ij}^{\oplus}\,\vec{e}_{z}\cdot \vec{n}_0 - H_{ij}^{\oplus}\,|\vec{e}_{z}\cdot \vec{n}_0|\Big)v_{i}\, u_{j}\, \mathrm{ds}\bigg),
\end{split}
\end{equation}
while the boundary contribution to $L$ is (cf.\,Eq.\,\ref{linearSource})
\begin{equation}
\begin{split}
L_i^{\text{BC}}(v_i) = -\bigg(&\int_{ \mathcal{B}_\mathrm{d}}\,\dfrac{1}{2}\Big(g_{\text{flux},x,i}^{\ominus}\,\vec{e}_{x}\cdot \vec{n}_0 - g_{i}^{\ominus}\,|\vec{e}_{x}\cdot \vec{n}_0|\Big)v_{i}\, \mathrm{ds} \\
+&\int_{ \mathcal{B}_\mathrm{d}}\,\dfrac{1}{2}\Big(g_{\text{flux},y,i}^{\ominus}\,\vec{e}_{y}\cdot \vec{n}_0 - g_{i}^{\ominus}\,|\vec{e}_{y}\cdot \vec{n}_0|\Big)v_{i}\, \mathrm{ds} \\
+&\int_{ \mathcal{B}_\mathrm{d}}\,\dfrac{1}{2}\Big(g_{\text{flux},z,i}^{\ominus}\,\vec{e}_{z}\cdot \vec{n}_0 - g_{i}^{\ominus}\,|\vec{e}_{z}\cdot \vec{n}_0|\Big)v_{i}\, \mathrm{ds}\bigg).
\label{eqLbc}
\end{split}
\end{equation}
Eq.\,\ref{a_ext} already accounts for the terms in Eq.\,\ref{eqFlux} that correspond to the inside of the exterior facet.

\paragraph{Reflective boundary}  
Due to the rotational invariance of the spherical harmonics, the reflected moment corresponding to $I_{j}$, $1\leq j\leq P$, can be expressed as:
\begin{equation}
I_{j'} = \sum_{j=1}^{P}\alpha_{j'j}\,I_{j},
\end{equation}
where $\alpha$ is matrix depending on $\vec{n}_0$.\footnote{In particular, if $\vec{n}_0 = \vec{e}_z$, $\alpha_{ij} = (-1)^{l+m}I_{j}$. Simple relations are also obtained if $\vec{n}_0 = \vec{e}_x$ or $\vec{n}_0 = \vec{e}_y$ \cite{McClarren_dissertation}.}
Hence, for all $1\leq i\leq P$,:
\begin{equation}
\begin{split}
a_i^{\text{BC},\mathrm{r}}(u,v_i) = \sum_{j = 1}^{P} \bigg(&\int_{ \mathcal{B}_\mathrm{r}}\,\dfrac{1}{2}\Big(A_{x,ij}\,\vec{e}_{x}\cdot \vec{n}_0 - M_{x,ij}\,|\vec{e}_{x}\cdot \vec{n}_0|\Big)v_{i}\, \alpha_{ij}(\vec{n}_0) u_{j}\, \mathrm{ds} \\
+&\int_{ \mathcal{B}_\mathrm{r}}\,\dfrac{1}{2}\Big(A_{y,ij}\,\vec{e}_{y}\cdot \vec{n}_0 - M_{y,ij}\,|\vec{e}_{y}\cdot \vec{n}_0|\Big)v_{i}\, \alpha_{ij}(\vec{n}_0) u_{j}\, \mathrm{ds} \\
+&\int_{ \mathcal{B}_\mathrm{r}}\,\dfrac{1}{2}\Big(A_{z,ij}\,\vec{e}_{z}\cdot \vec{n}_0 - M_{z,ij}\,|\vec{e}_{z}\cdot \vec{n}_0|\Big)v_{i}\, \alpha_{ij}(\vec{n}_0) u_{j}\, \mathrm{ds}\bigg).
\label{eqReflbc}
\end{split}
\end{equation}

\subsection{Implementation}
%
%
%
\label{mooseCode}
To generate numerical solutions for Eqs.~\ref{eq0} and \ref{eq01}, a code has been implemented in Rattlesnake, the transport solver of the the Idaho National Laboratory (INL), based on the Multiphysics Object Oriented Simulation Environment (MOOSE) framework \cite{MOOSE}.
Nonlinear solves are performed using the Jacobian Free Newton Krylov (JFNK) method, and the PETSc \cite{PETSc} restarted generalized minimal residual (GMRES) solver for the linear solves. In this method, the Jacobian is never explicitly formed but its action is computed with two nonlinear residual evaluations. All the results from this code are obtained using the first order LAGRANGE elements from libMesh \cite{libMeshPaper}. The meshes are generated using {\tt gmsh} \cite{Gmsh} and the results are visualized with VisIt \cite{VisIt}.  Several convergence tests were performed to verify the spatial and temporal accuracy of the code.

The linear system for $I$ in Eq.\,\ref{eq1} can be ill-conditioned in streaming regimes.  Specifically, $\sigma_{\mathrm{t}}^{*} \to  \frac{1}{c\Delta t}$ when $\sigma_{\mathrm{t}} \to 0$.  Hence when $\sigma_{\mathrm{t}}$ is small and $\Delta t$ is large, the system is dominated by the streaming operator $A_{x}\frac{\partial I}{\partial x} + A_{y}\frac{\partial I}{\partial y} + A_{z}\frac{\partial I}{\partial z}$, which is singular and not diagonally dominant. The loss of diagonal dominance makes most iterative schemes (Jacobi, Gauss-Seidel, SOR, etc.) unstable.  To our knowledge, there does not exist a universally effective preconditioner for the P$_{N}$ equations in the streaming limit, though some multigrid in angle preconditioners have been studied in the past for the even-parity form of the P$_{N}$ equations \cite{PN_precond_Oliveira}.  For the results in this paper, we have used the built-in algebraic multigrid (AMG) preconditioners in PETSc.

\section{Study of the filter}
\label{Sec4}

In this section, we discuss the selection of filter parameters.  We then investigate how the filter affects (i) the convergence of the iterative solver for the fully discretized system and (ii) the convergence of the angular discretization as $N \to \infty$.
\subsection{Filtering strategy}
\label{SecFilteringStrategy}
In this subsection, we discuss the strategy for selecting the location, type, and strength of the filter.
%


The major drawback of the filter is that $\sigf$ must be tuned by the user for each individual problem.  Unfortunately, the numerical solution can be very sensitive to the value of $\sigf$, especially for small values of $N$.  The choice of filter strength is a trade-off between removing unphysical oscillations and excessive damping of the solution.  Since the appropriate balance may be different in different parts of the spatial domain, it is often advantageous to allow $\sigf$ to vary in space.  Often a basic understanding of radiation transport can help guide the strategy for setting $\sigf$ without the need for extensive knowledge of the solution beforehand.  When more information is needed, a relatively coarse simulation (in space and time) may be used as a proxy.  This is one of the main benefits of using the consistent formulation in Eq.\,\ref{eq11}:  the value $\sigf$ does not need to be recomputed when the space-time mesh is refined.

In our experience, we have found the following to be good practices for setting the filter strength.
\begin{itemize}
	\item {\bf Location.} Run a calculation with no filter and find local regions at which $I_0^0$ becomes negative.  Activate the filter in these `negative' regions as well as in upstream regions of comparable sizes. For the other parts of the problem, the filter can typically be set to zero or to a much smaller value. If the problem is uniform, then activate the filter everywhere.
	\item {\bf Filter type.} Set the order of the filter to match the expected regularity (with respect to angle) of the transport solution.%
	\footnote{See Sec.\,\ref{SecErrorEstimates} for a more precise statement of the regularity.}  
	If unsure, it is better to underestimate the regularity.  Lower order filters are typically more robust \addition{because they damp the lower order moments more strongly.\footnote{\addition{In particular, an unfiltered calculation can be seen as a filtered calculation of order $\infty$.}}} For the most difficult problems, we have found that the second-order Lanczos filter works well.
	This is the filter used by default throughout this paper.
	\item {\bf Filter strength.} \addition{Using a coarse mesh, determine $N_0$ which yields an acceptable\footnote{\addition{As $N\to\infty$, the numerical solution converges to the analytical solution so there exists an integer $N_0$ such that the numerical solution is subjectively good enough. In practice, $N_0$ can be chosen such that the unfiltered $I_0^0$ is non-negative.}} unfiltered solution.}  A good scaling is usually obtained by setting $\sigma_{\text{f}}(\vr) \approx \sigt(\vr)/f(1,N_0)$ in the previously determined regions.
	\addition{Another option is to tune the filter strength empirically.}
\end{itemize}

These guidelines are quite broad but they usually are precise enough to determine a suitable $\sigf$. The relative freedom that is left to the user is also an advantage since the extent to which the negativity and oscillations should be reduced can vary from one application to another.

\subsection{Effects of the filter on the iterative solver}
\label{FilterEffect}
We \
consider the effect of the filter on the iteration count for the full nonlinear system when solving the Crooked Pipe problem.\
A full description of this problem can be found in Section \ref{Sec5} (see Fig.\,\ref{fig2} for the layout), and numerical solutions are presented later in Section \ref{Sec5}. 

\begin{figure}[H]
	\centering
	\begin{subfigure}[b]{0.45\textwidth}
		\includegraphics[width=\textwidth,height=0.8\textwidth]{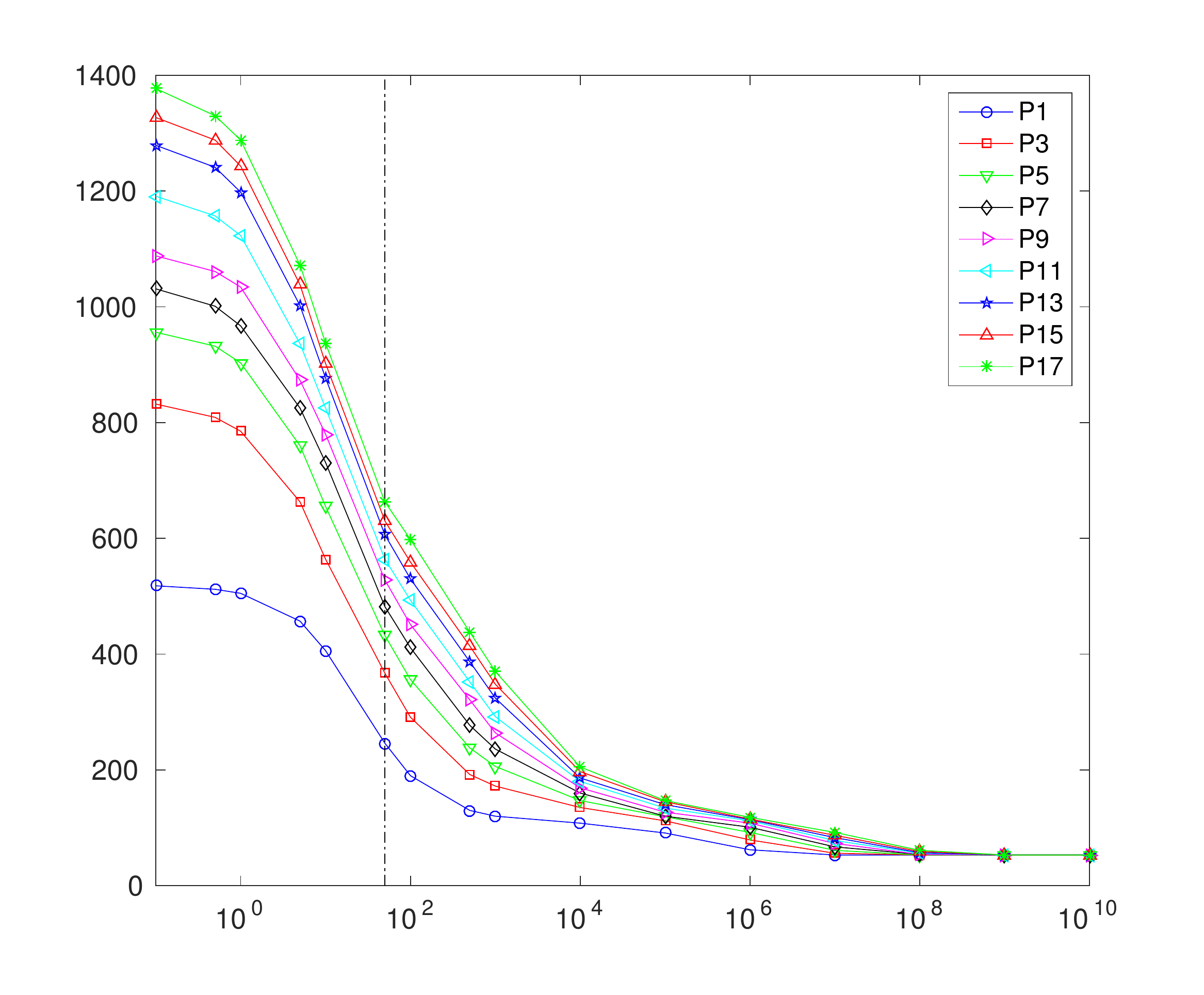}
		\caption{Uniform filtering}
	\end{subfigure}%
	\begin{subfigure}[b]{0.45\textwidth}
		\includegraphics[width=\textwidth,height=0.8\textwidth]{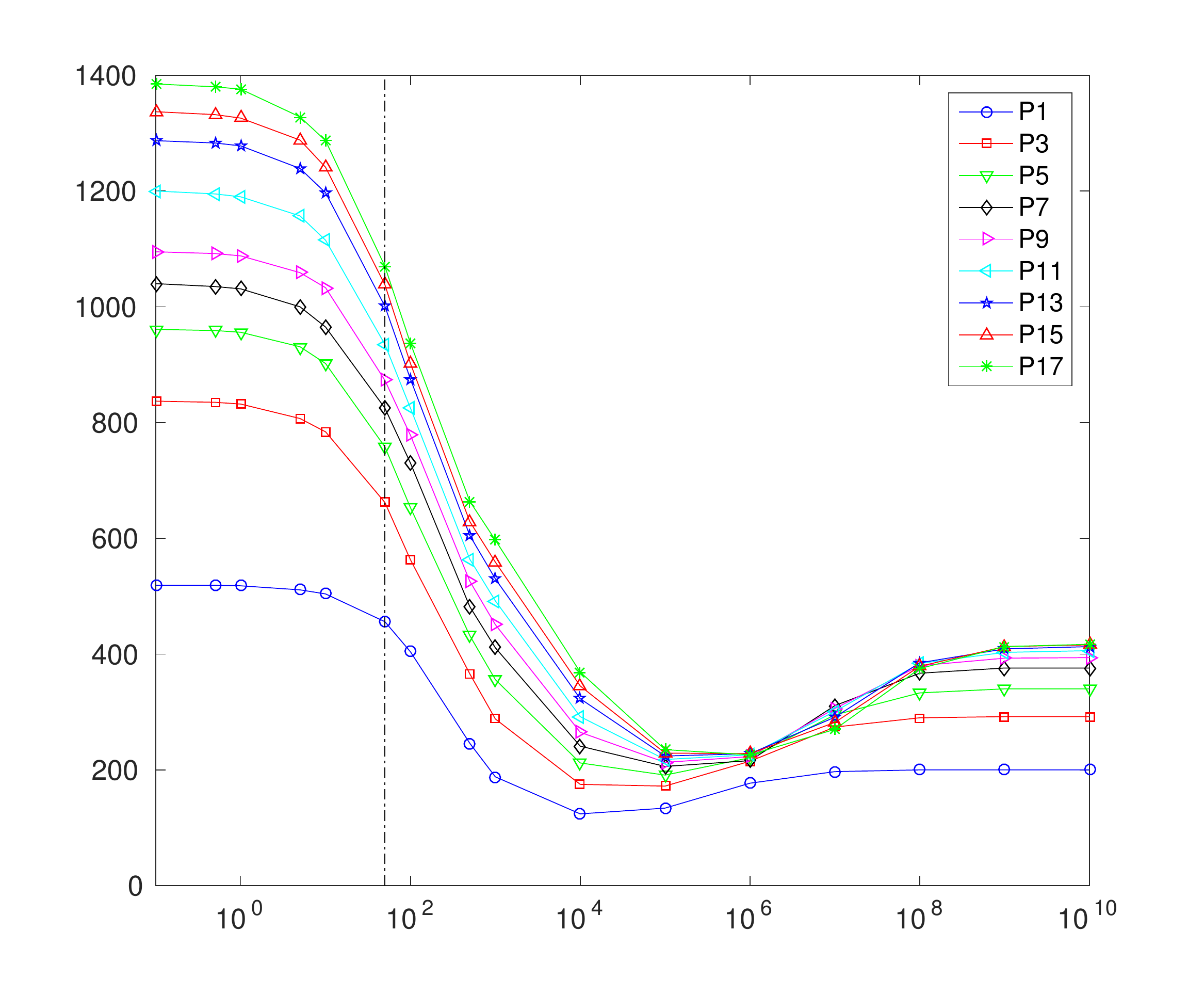}
		\caption{Local filtering}
	\end{subfigure}
	\caption{Iteration count for the first time step as a function of $N$ and the filter strength $\sigf$ (in cm$^{-1}$), using the Lanczos filter.  As a reference, the value of $\sigf$ for this test problem was in practice chosen to be 50 cm$^{-1}$ (vertical line).  The value of $\sigf$ for the local filter designates the maximum value; see Fig.\,\ref{fig:localFiltering} for a complete description.}
	\label{fig:filterImpactOnCV}
\end{figure}

In Fig.\,\ref{fig:filterImpactOnCV}, the total number of GMRES iterations are displayed for the first time step, which is typically the most expensive. 
For the uniform filter, the number of iterations decreases monotonically as $\sigf$ increases to a fixed number that is independent of $N$.  For the local filter, the iterations decrease initially and increase to a fixed value that is different for each $N$. (Note however, that this increase occurs well beyond any practical value of $\sigf$.)   The difference in performance between the uniform and local strategies is due to the fact that the local filter introduces an artificial discontinuity in the effective material cross-section.     In both strategies, the improvement in performance is noteworthy.  Indeed, the number of iterations for the practical value of $\sigf$ decreases by more than one-half when compared to the unfiltered case for uniform filtering and by more than 20\% for the locally filtered P$_N$ with $N>1$. \\

\addition{The decrease in the number of iterations for small values of $\sigf$ as well as the convergence to a constant number for $\sigf\to\infty$ can be predicted on a pure transport problem using GMRES convergence properties. This is because the filtering operator only adds a diagonal contribution to the global matrix which tends to gather the eigenvalues into $N+1$ clusters as $\sigf\to\infty$, $N$ of which having a relative radius going to zero in that limit.
	Detailed derivations were removed from this work for conciseness but can be found in \cite{LaboureThesis}.}

\subsection{Comparison to error estimates}

\label{SecErrorEstimates}
Frank, Hauck and Kuepper \cite{FranckHauck2014} have derived error estimates for the convergence of filtered P$_N$ for the case of pure transport.  Here we compare these estimates to numerical results for smooth and non-smooth solutions of thermal radiative transfer with non-linear material properties.  Define the angular error
\begin{equation}
E_{N} = ||\hat{\mathcal{I}}_{N} - \mathcal{I}||_{L^{2}} = \bigg( \sum_{\ell=0}^{\infty}\sum_{m=-\ell}^{\ell}\int_{\mathcal{D}}\,\Big((\hat{\mathcal{I}}_{N})_{\ell}^{m} - \mathcal{I}_{\ell}^{m}\Big)^{2}  \mathrm{d}x\bigg)^{1/2},
\end{equation}
where the expansion coefficients of $\hat{\mathcal{I}}_{N}$ solve the (time continuous) FP$_N$ equations and we have added the subscript $N$ to $\hat{\mathcal{I}}$ to emphasize the dependence on $N$.  Based on  \cite{FranckHauck2014}, we expect  
\begin{equation}
E_{N} = \mathcal{O} \left(N^{-\min\{k,\alpha\}}\right),
\label{eqErrEstimates}
\end{equation}
where  $k$ is the order of convergence in the unfiltered case, $\alpha$ is the order of the filter
and the implied constant in Eq.\,\ref{eqErrEstimates} depends on $\mathcal{I}$ and the time $t$. The Lanczos and Spherical Spline filter orders are two and four, respectively. 

As a test problem, we use the smooth Marshak Wave \cite{Lowrie2004}.  This problem is defined on a slab geometry, which implies that $I$ only depends on $x \in [0,1]$ and $t$.  It assumes a purely absorbing medium with cross-section $\sigma_{\mathrm{t}}=\sigma_{\mathrm{a}}=(ac)^{-3/4}\, T^{-3}$. The material heat capacity $C_v$ is set to $a^{1/4}c^{-3/4}$.%
\footnote{In the original paper, the equations solved can be obtained by setting $a = c = C_v = 1$. Here we prefer to keep the physical constants unchanged and use a different scaling, which leads to slightly different expressions for the cross-sections, the heat capacity, the time step and the temperature. We are however solving the same equations.} The initial conditions are 
\begin{align}
I_{\ell}^{m}(x,0) &= \frac{\delta_{\ell, 0}}{\sqrt{4\pi}}\Big(\mathcal{I}_{l} + (\mathcal{I}_{r}-\mathcal{I}_{l})\frac{1+\tanh\big(50(x-0.25)\big)}{2}\Big) ,\\
T(x,0) &= \Big(\dfrac{\sqrt{4\pi}}{ac}I_{0}^{0}(x,0)\Big)^{1/4},
\end{align}
where $\mathcal{I}_{l}=4$, $\mathcal{I}_{r}=0.004$. 
We use Dirichlet boundary conditions (see Sec.\,\ref{BCsec}) at both boundaries: $g=\mathcal{I}_{l}$ at $x=0$ and $g=\mathcal{I}_{r}$ at $x=1$. 
We use $200$ uniform cells of width $\Delta x$ = 0.005.  The final time is $t_{\max} = \Delta t$ = 0.005/$c$ and the filter strength is $\sigf= 100$. 

To test both aspects of \eqref{eqErrEstimates}, we consider two problems. In the first one, $\mathcal{Q}=0$; in the second, we add a non-smooth, volumetric source that is constant in $x$ and $t$ and a hat function in $\mu$.
\begin{equation}
\mathcal{Q}(x,\mu,t) = \begin{cases}
20\big|(|\mu|-0.5)\big|, & 0\leq |\mu|\leq 0.5 , \\
0, & 0.5 \leq |\mu| \leq 1.
\end{cases}
\end{equation} 
Thus the angular derivative is not continuous.
To estimate the error $E_N$, we use $\hat{\mathcal{I}}_{99}$ and $\hat{\mathcal{I}}_{199}$, respectively, to approximate $\mathcal{I}$ in the smooth and non-smooth cases.\footnote{In the non-smooth case, the reference solution must be more refined in order to see a more saturated convergence rate.
}



In Tables \ref{smooth1}--\ref{smooth3} we show numerical values of $E_{N_{i}}$ and the convergence rate
\begin{equation}
r_{i} = - \dfrac{\log (E_{N_{i}} / E_{N_{i+1}}) }{\log (N_{i} / N_{i+1}) },
\end{equation}
for several different filters in the smooth case.  As expected, the order of convergence is close to the order of the filter.
\vspace{0.5cm}\\
\begin{minipage}{\textwidth}
	\hfill
	\begin{minipage}[t]{0.3\textwidth}
		\centering
		\begin{tabular}{ | c | c | c |}
			\hline
			$N$ &  $E_{N}$ & $r$   \\[2ex] \hline
			1	&	1.29E-06	&	4.35	\\
			3	&	1.08E-08	&	7.59	\\
			7	&	1.74E-11	&	5.04	\\
			15	&	3.75E-13	&	3.78	\\
			29	&	3.10E-14	&	0.68	\\
			49	&	2.17E-14	&	NA	\\
			99	&	Reference	&	NA	\\
			\hline
		\end{tabular}
		\captionof{table}{Unfiltered (smooth)}
		\label{smooth1}
	\end{minipage}
	\quad
	\begin{minipage}[t]{0.3\textwidth}
		\centering
		\begin{tabular}{ | c | c | c |}
			\hline
			$N$ &  $E_{N}$ & $r$   \\[2ex] \hline
			1	&	1.59E-06	&	2.01	\\
			3	&	1.75E-07	&	1.64	\\
			7	&	4.36E-08	&	1.83	\\
			15	&	1.08E-08	&	1.97	\\
			29	&	2.95E-09	&	2.15	\\
			49	&	9.56E-10	&	NA	\\
			99	&	Reference	&	NA	\\
			\hline
		\end{tabular}
		\captionof{table}{Lanczos (smooth)}
	\end{minipage}
	\quad
	\begin{minipage}[t]{0.3\textwidth}
		\centering
		\begin{tabular}{ | c | c | c |}
			\hline
			$N$ &  $E_{N}$ & $r$   \\[2ex] \hline
			1	&	1.78E-06	&	2.84	\\
			3	&	7.89E-08	&	3.25	\\
			7	&	5.00E-09	&	3.64	\\
			15	&	3.13E-10	&	3.82	\\
			29	&	2.53E-11	&	3.91	\\
			49	&	3.26E-12	&	NA	\\
			99	&	Reference	&	NA	\\
			\hline
		\end{tabular}
		\captionof{table}{SSpline (smooth)}
		\label{smooth3}
	\end{minipage}
\end{minipage}

\

In Tables \ref{nonsmooth1}--\ref{nonsmooth3} we show the results in the non-smooth case. We observe that the order of convergence is not affected by the order of the filter.  This is as expected, since $k < \alpha$. Because lower-order filters are more robust, it is generally best to choose $\alpha$ no less than $k$, but as close to $k$ as possible. \addition{Results confirming Eq.~\ref{eqErrEstimates} are similarly obtained for the exponential filters of an arbitrary order (which are introduced in \cite{FranckHauck2014}).}
\vspace{0.5cm}\\
\noindent
\begin{minipage}{\textwidth}
	\hfill
	\begin{minipage}[t]{0.30\textwidth}
		\centering
		\begin{tabular}{ | c | c | c |}
			\hline
			$N$ &  $\quad E_{N}\quad$ & $r$   \\[2ex] \hline
			1	&	1.60E-02	&	0.23	\\
			3	&	1.23E-02	&	1.40	\\
			7	&	3.76E-03	&	1.43	\\
			15	&	1.27E-03	&	1.50	\\
			29	&	4.73E-04	&	1.63	\\
			49	&	2.01E-04	&	1.57	\\
			69	&	1.18E-04	&	1.11	\\
			89	&	8.87E-05	&	1.65	\\
			109	&	6.35E-05	&	NA	\\
			199	&	Reference	&	NA	\\
			\hline
		\end{tabular}
		\captionof{table}{Unfiltered (non-smooth)}
		\label{nonsmooth1}
	\end{minipage}
	\quad
	\begin{minipage}[t]{0.3\textwidth}
		\centering
		\begin{tabular}{ | c | c | c |}
			\hline
			$N$ &  $E_{N}$ & $r$   \\[2ex] \hline
			1	&	1.60E-02	&	0.23	\\
			3	&	1.23E-02	&	1.40	\\
			7	&	3.78E-03	&	1.43	\\
			15	&	1.27E-03	&	1.49	\\
			29	&	4.75E-04	&	1.63	\\
			49	&	2.02E-04	&	1.56	\\
			69	&	1.19E-04	&	1.12	\\
			89	&	8.93E-05	&	1.64	\\
			109	&	6.40E-05	&	NA	\\
			199	&	Reference	&	NA	\\
			\hline
		\end{tabular}
		\captionof{table}{Lanczos (non-smooth)}
	\end{minipage}
	\quad
	\begin{minipage}[t]{0.3\textwidth}
		\centering
		\begin{tabular}{ | c | c | c |}
			\hline
			$N$ &  $E_{N}$ & $r$   \\[2ex] \hline
			1	&	1.60E-02	&	0.23	\\
			3	&	1.23E-02	&	1.38	\\
			7	&	3.84E-03	&	1.45	\\
			15	&	1.27E-03	&	1.48	\\
			29	&	4.79E-04	&	1.61	\\
			49	&	2.06E-04	&	1.55	\\
			69	&	1.21E-04	&	1.14	\\
			89	&	9.06E-05	&	1.63	\\
			109	&	6.51E-05	&	NA	\\
			199	&	Reference	&	NA	\\
			\hline
		\end{tabular}
		\captionof{table}{SSpline (non-smooth)}
		\label{nonsmooth3}
	\end{minipage}
\end{minipage}
\section{Numerical Solutions for Crooked Pipe and Comparison with IMC}
\label{Sec5}

\begin{figure}[H]
	\centering
	\hspace*{-0.5cm}
	\includegraphics[width=1\textwidth]{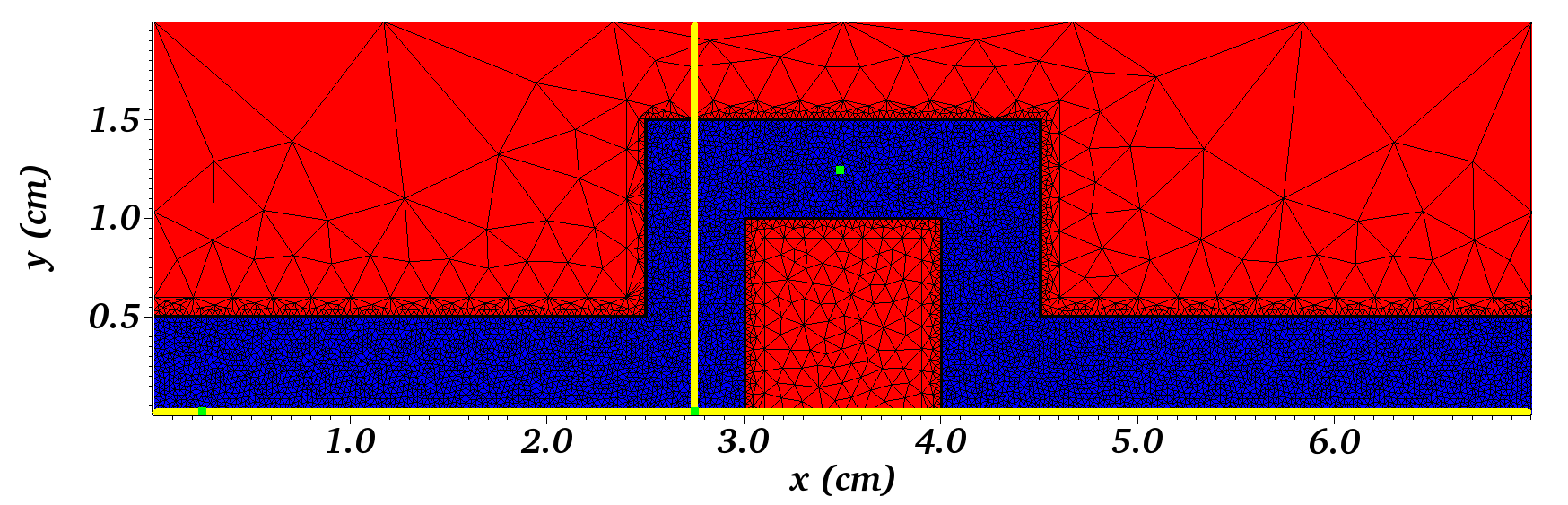}
	\caption{Mesh for the Crooked Pipe test problem. In the thin regions (shown in blue), $\sigma_{\mathrm{t}}=\sigma_{\mathrm{a}}=20$ m$^{-1}$ and $C_{v}=4.3\times 10^{4}$ J/m$^{4}$/K. In the thick regions (shown in red), each of these constants is factor of $1000$ greater. The two straight lines (in yellow) are $y = 0$ and $x=2.75$ cm; the three points (in green) are $(x_{1},y_{1})=$ (0.25\,cm,\,0), $(x_{2},y_{2})=$ (2.75\,cm,\,0) and $(x_{3},y_{3})=$ (3.5\,cm,\,1.25\,cm). The interface between thick and thin regions is refined so that there are several cells per mean free path. (The first layer of cells has a width of 0.005 cm.) The entire mesh contains 20,106 triangular elements.}
	
	\label{fig2}
\end{figure}

In this section, we study a variation%
\footnote{The original Crooked Pipe problem has a cylindrical geometry; here we use Cartesian coordinates.}
of the Crooked Pipe benchmark \cite{CrookedPipe}.  In this problem, there are two purely absorbing materials in a two-dimensional, Cartesian domain that is 7 cm $\times$ 2 cm, respectively, in the $x$ and $y$ directions (as shown in Fig.\,\ref{fig2}), with the origin located at the bottom left corner.  There is no $z$-dependence.  The location of the two materials is shown in Fig.\,\ref{fig2}. In the thin one, $\sigma_{\mathrm{a}} = 20$ m$ ^{-1}$ and $C_{v}=4.3\times 10^{4}$ J/m$^{3}$/K; in the thick one, $\sigma_{\mathrm{a}} = 2\times 10^{4}$ m$ ^{-1}$ and $C_{v}=4.3\times 10^{7}$ J/m$^{3}$/K. 

On the left boundary, we apply an isotropic incoming source (see Eq.\,\ref{Dirichlet}): 
\begin{equation}
g=\frac{ac}{\sqrt{4\pi}}T^{4}_{L}, \quad T_{L}=0.3~ \rm{keV},
\end{equation}
at $x=0$ for $0\leq y\leq 0.5$ cm---that is, only along the thin region of the left boundary. We also apply a 0.05 keV source on the thin region of the right boundary to keep particles from leaking out of the domain there.  A reflective boundary condition is imposed on the bottom boundary and open boundaries are imposed everywhere else. The initial temperature is set to $T_{0}=0.05$  keV, and the expansion coefficients of the initial intensity are
\begin{equation}
I_{\ell}^{m}(x,0) = \frac{acT_{0}^{4}}{\sqrt{4\pi}} \delta_{\ell, 0}.
\end{equation}

As explained in Section \ref{SecLumping}, we lump the mass matrix for the collision terms in order to increase robustness. The time step is set to 0.05 ns using a BDF-2 time-discretization scheme.%
\footnote{The difference with the Backward-Euler scheme was barely noticeable, suggesting that the temporal error is not dominant with this time step. Increasing the time step to 0.1 ns also had a negligible impact.}

\subsection{Comparison with IMC: Simplified Problem}

\begin{figure}[H]
	\begin{subfigure}[b]{0.32\textwidth}
		\includegraphics[width=\textwidth]{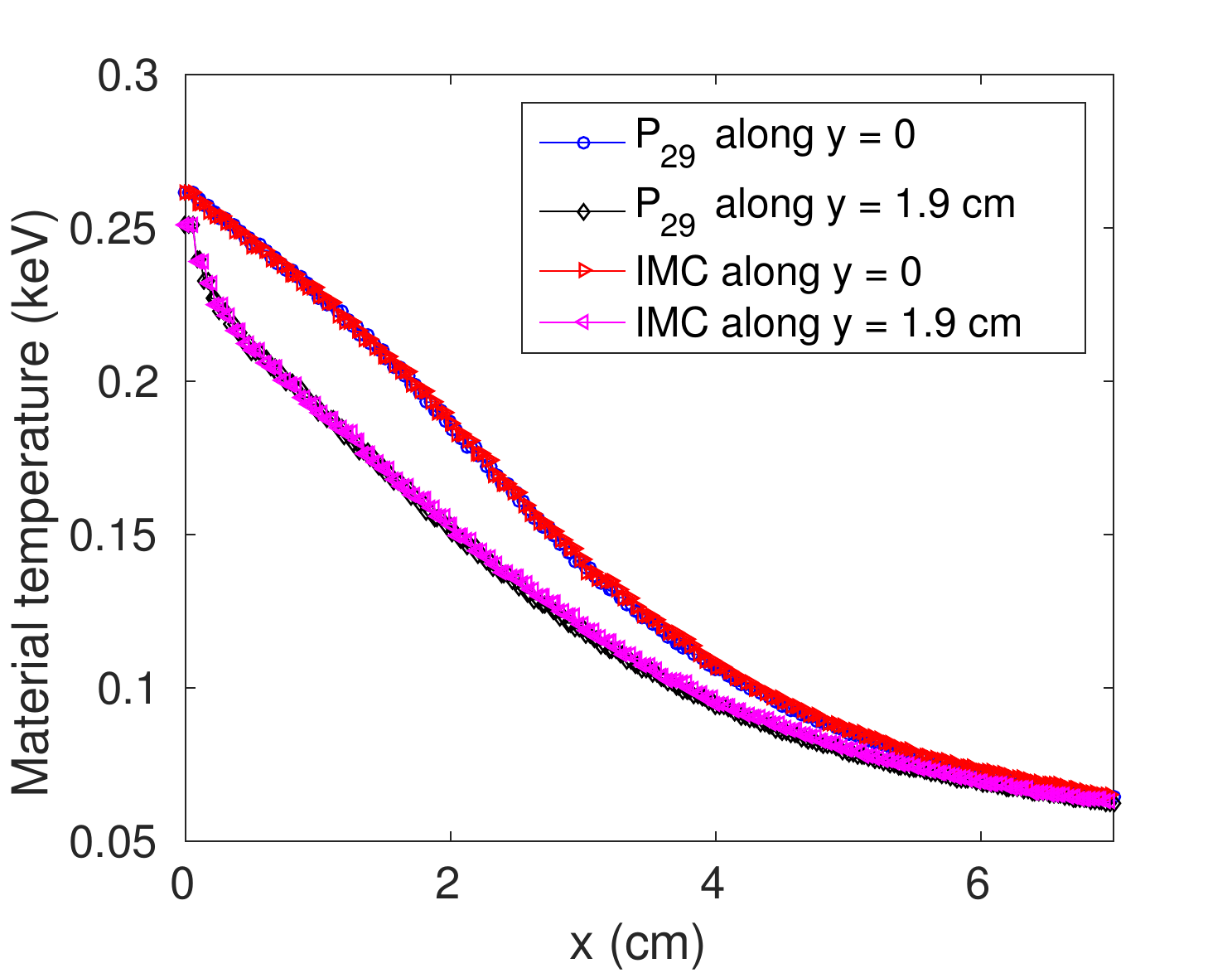}\\
		\caption{$T$ at $t=1$ ns.}
		\label{fig:PN_IMC_1}
	\end{subfigure}%
	\begin{subfigure}[b]{0.32\textwidth}
		\includegraphics[width=\textwidth]{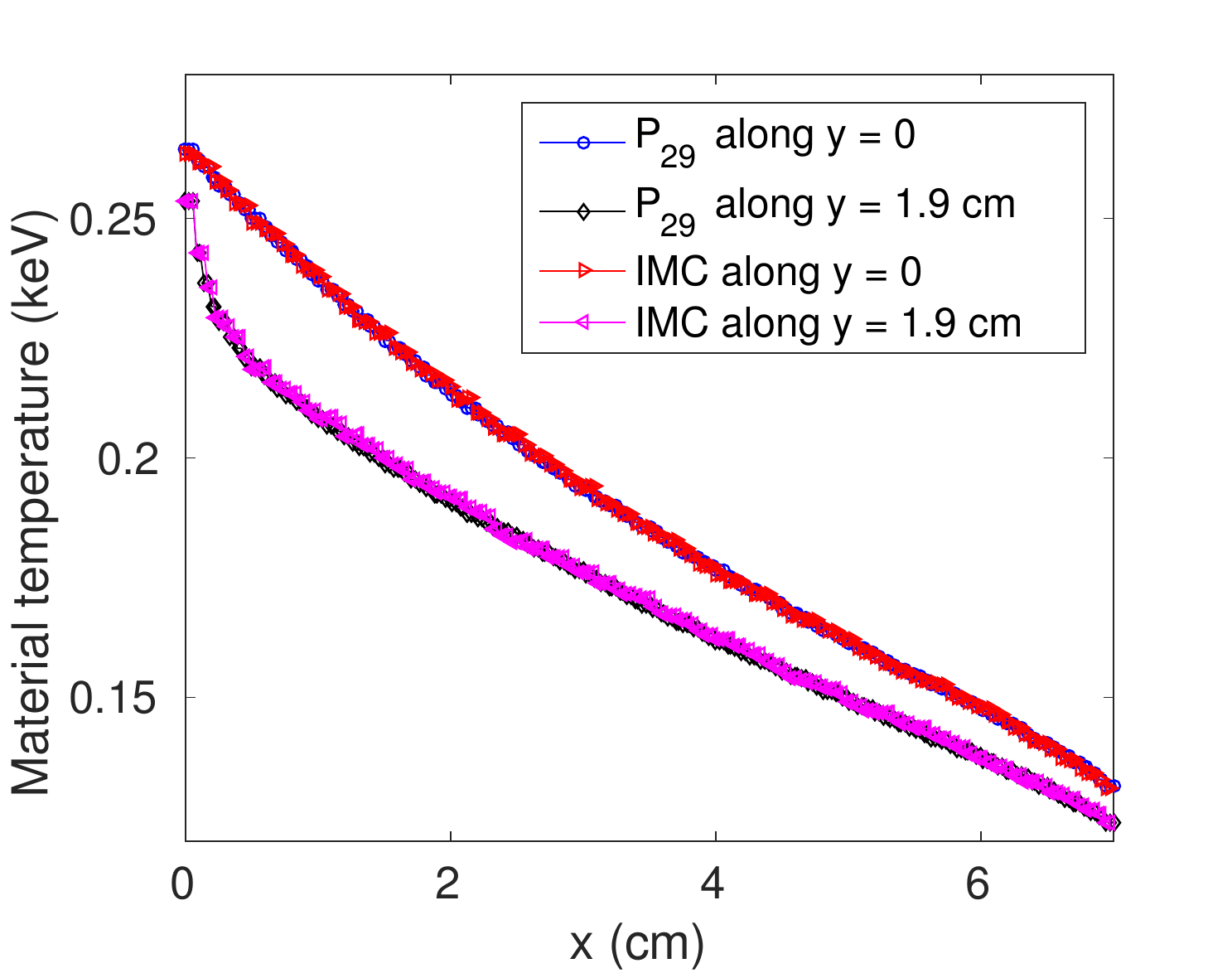}\\
		\caption{$T$ at $t=10$ ns.}
		\label{fig:PN_IMC_2}
	\end{subfigure}
	\begin{subfigure}[b]{0.32\textwidth}
		\includegraphics[width=\textwidth]{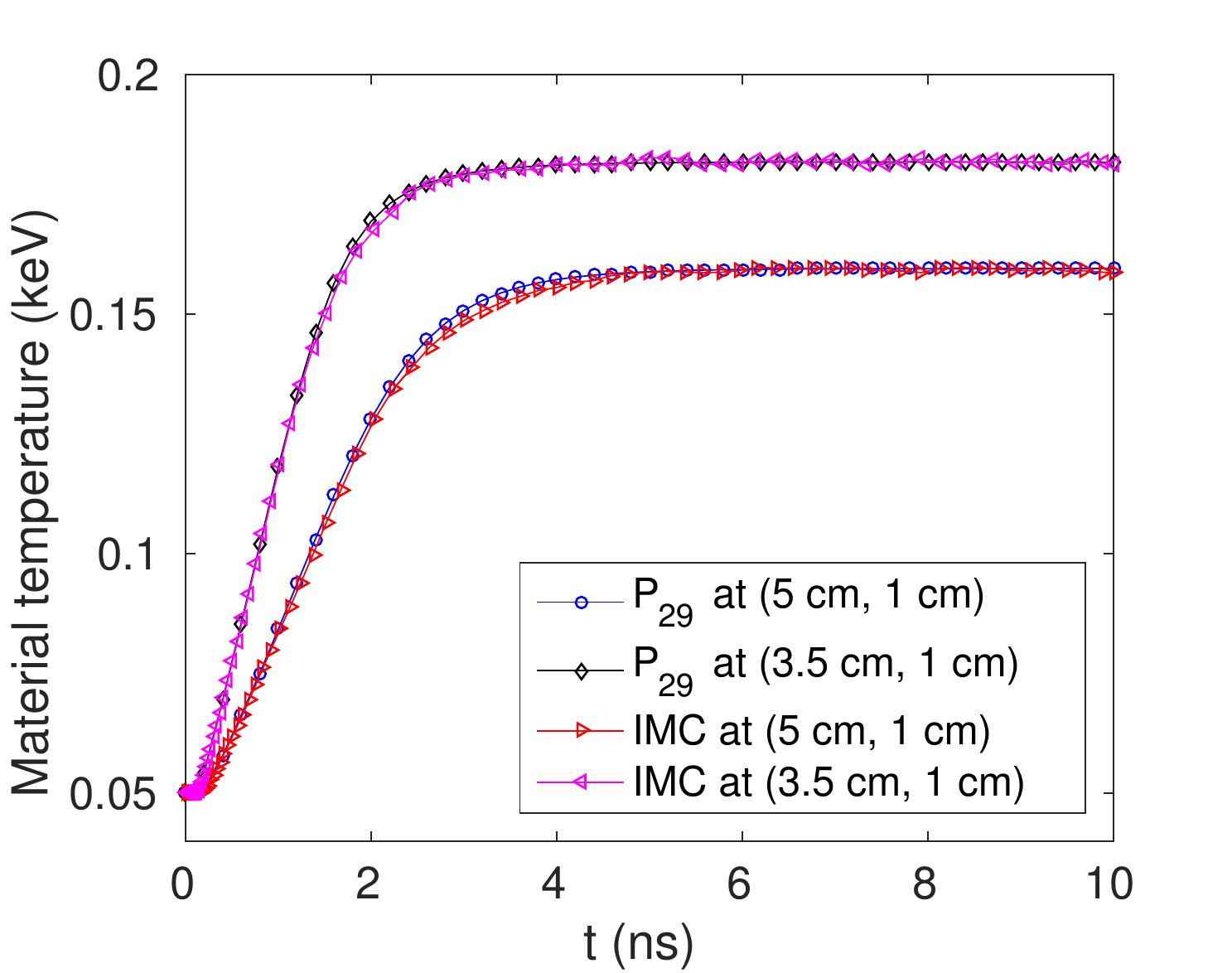}\\
		\caption{$T$ at two spatial points.}
		\label{fig:PN_IMC_3}
	\end{subfigure}
	\captionsetup{margin=0.5cm}
	\caption{Temperature profile along $y=0$, $y = 1.9$ cm as a function of $x$ and at $(x,y) =$ (3.5 cm, 1 cm)  and $(x,y) =$ (5 cm, 1 cm) as a function of time. For convergence purposes, these results are obtained on the same geometry as Fig.\,\ref{fig2} except that the material properties are set to the thin region everywhere and that the source is applied on the entire left boundary. The mesh however was a uniform rectangular grid (100$\times$50 for the IMC, 112$\times$32 for the P$_{29}$). }
	\label{fig:PN_IMC}
\end{figure}

The sharp material interfaces and the absence of scattering in the Crooked Pipe make it very difficult to solve.  Furthermore, because $\sigma_{\rm{a}}$ in the thick region is very large, fully converging the solution requires a significant amount of computational resources. Thus, for verification purposes, we begin with a simpler test problem and compare it to a solution obtained from an IMC calculation. In this problem, $\sigma_{\mathrm{a}} = 20$ m$ ^{-1}$ everywhere and the source on the left is applied along the entire left boundary.  We verify that a P$_{29}$ solution agrees well with the IMC one; see Fig.\,\ref{fig:PN_IMC}. With this fact in mind, we use a P$_{39}$ solution with the spatial mesh shown in Fig.\,\ref{fig2} as the reference solution below.

\subsection{Filtering strategy}
For robustness, we use the Lanczos filter in all of the filtered calculations. Based on the guidelines detailed in the previous section, we consider three filtering strategies.

\begin{figure}[H]
	\includegraphics[width=\textwidth]{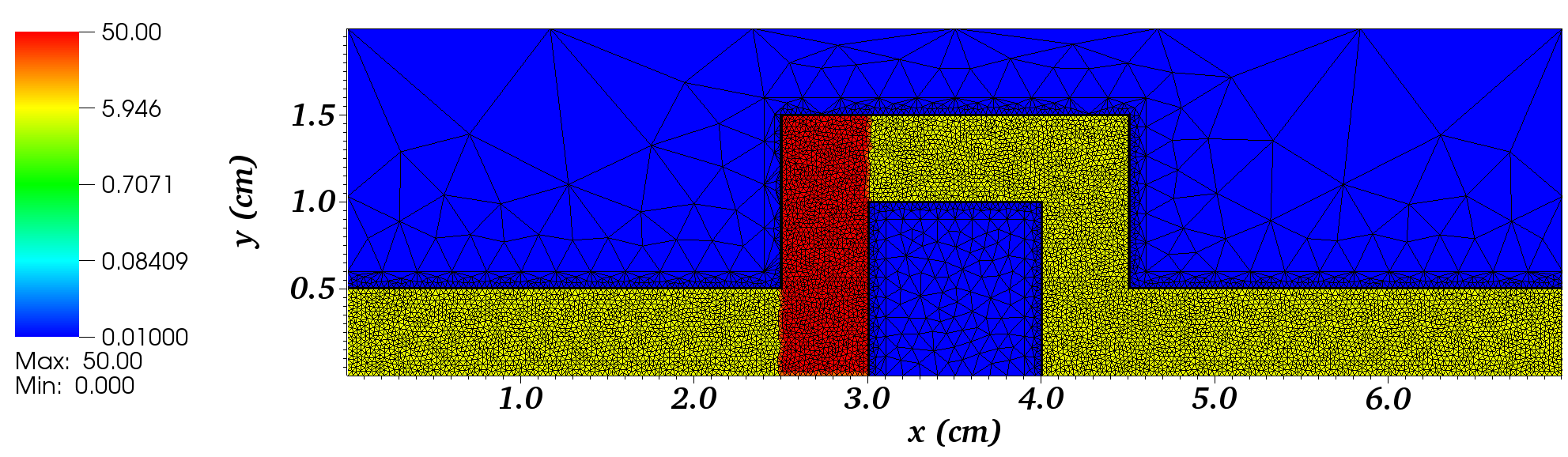}\\
	\caption{Value of $\sigf$ (in cm$^{-1}$) for the locally filtered calculations.}
	\label{fig:localFiltering}
\end{figure}

\begin{itemize}
	\item {\bf Unfiltered.} This is the original P$_{N}$ method, obtained by setting $\sigf =0$. 
	
	\item {\bf Uniformly filtered.} Here $\sigf$ is a fixed constant across the domain. Based on the discussion in Section \ref{SecFilteringStrategy} and given that the material temperature $T$ is virtually always above the initial temperature for $N = 7 $, we choose a value such that $\sigf\,f(1,N_0=7)$ is comparable to the cross-section in the thin part of the problem. Setting $\sigf = 5\times 10^{3}$ m$^{-1}$ gives $\sigf\,f(1,7) \approx 13$ m$^{-1}$ 
	. (Recall that $\sigma_{\mathrm{t}} = 20$ m$ ^{-1}$ in the thin region.)

	\item  {\bf Locally filtered.} The spatial profile of $\sigf$ in this case is provided by Fig.\,\ref{fig:localFiltering}.  Following the guidelines of  Section \ref{SecFilteringStrategy}, we set it to $5\times 10^{3}$ m$^{-1}$ after the first elbow of the pipe (where the radiation tends to become negative) as well as in an upstream region of comparable size. 
\end{itemize}

\subsection{Results}

In all simulations, radiation flows rapidly from the left boundary to the first elbow of the pipe. It is then absorbed and re-emitted by the material.  Isotropic re-emission allows for some of the radiation to change direction and propagate further down the pipe.  

In the following subsections, we present 2-D maps of the different solutions at a fixed time.  
We then examine these solutions in more detail: first along specified lines in space with time fixed  and then at fixed points in space over a given time interval. 
As expected, the locally filtered strategy generally produces the best solutions:  it maintains a positive scalar intensity without damping its profile too strongly.

\subsubsection{\addition{Scalar intensity} 2-D maps}

\begin{figure}[H]
	\centering
	\includegraphics[width=0.99\textwidth]{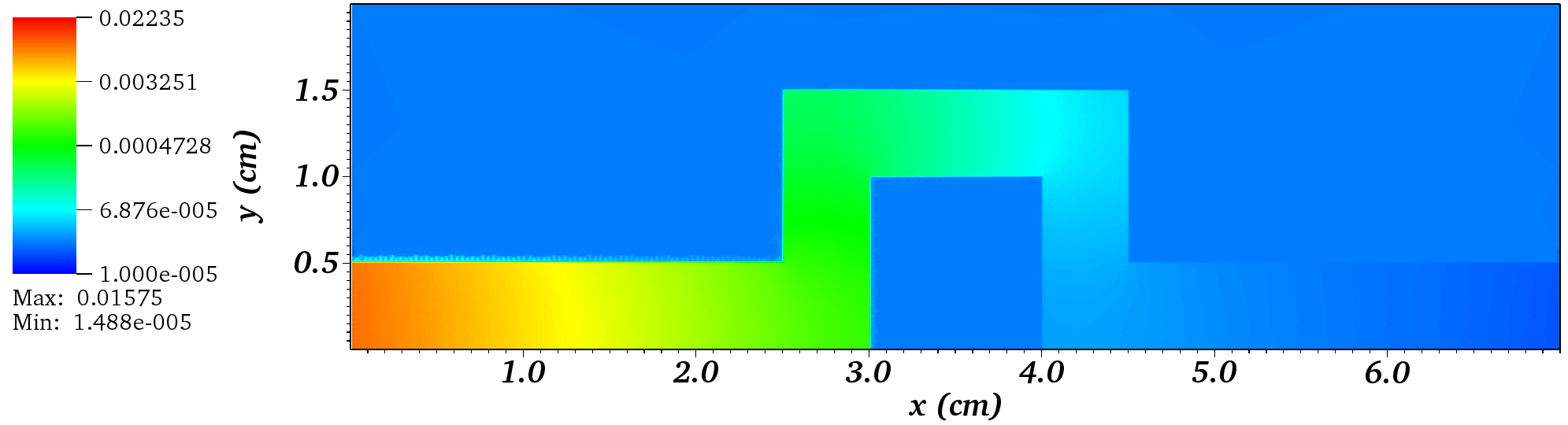}\\
	\includegraphics[width=0.99\textwidth]{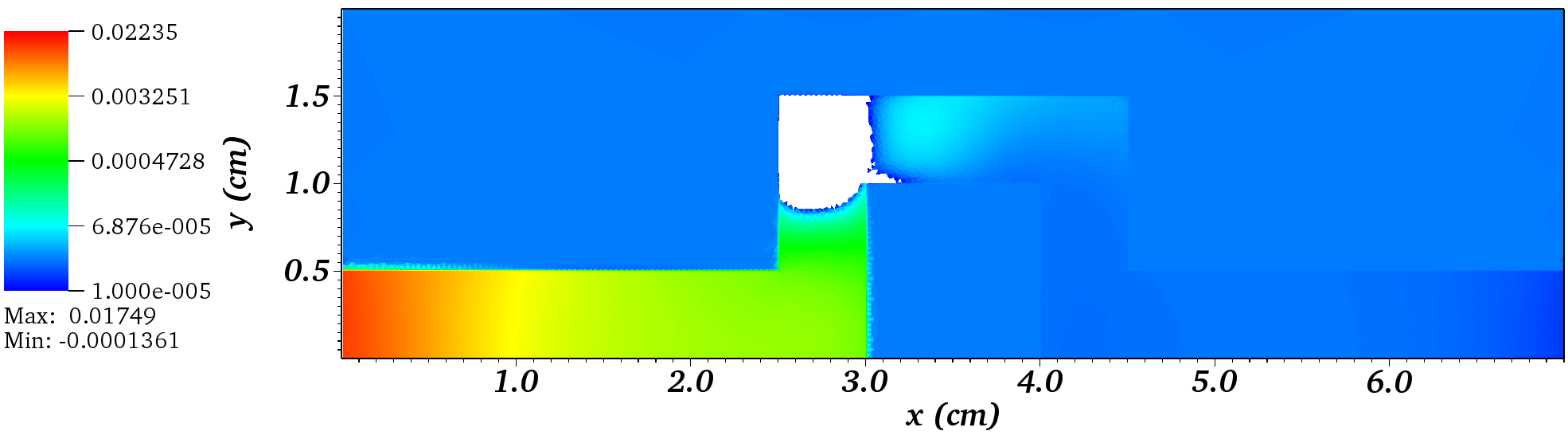}\\
	\includegraphics[width=0.99\textwidth]{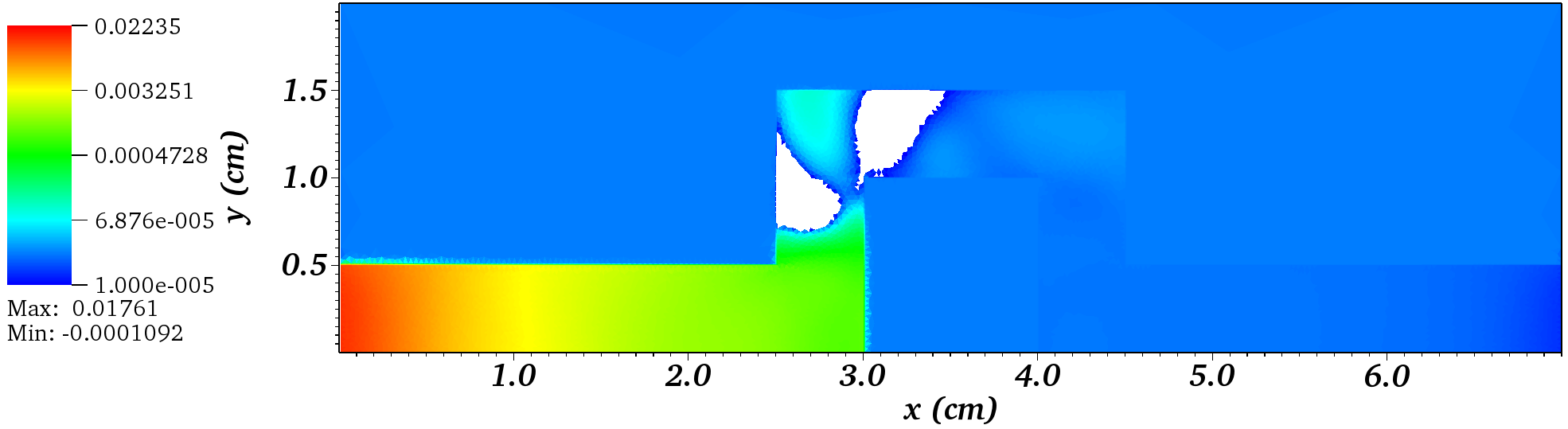}\\
	\includegraphics[width=0.99\textwidth]{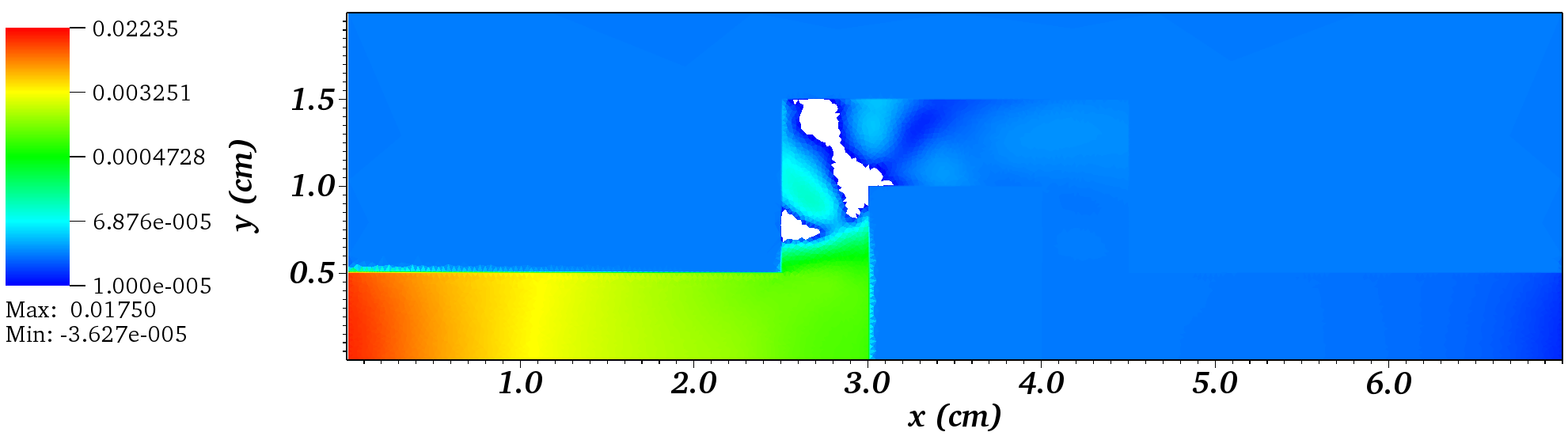}\\
	\includegraphics[width=0.99\textwidth]{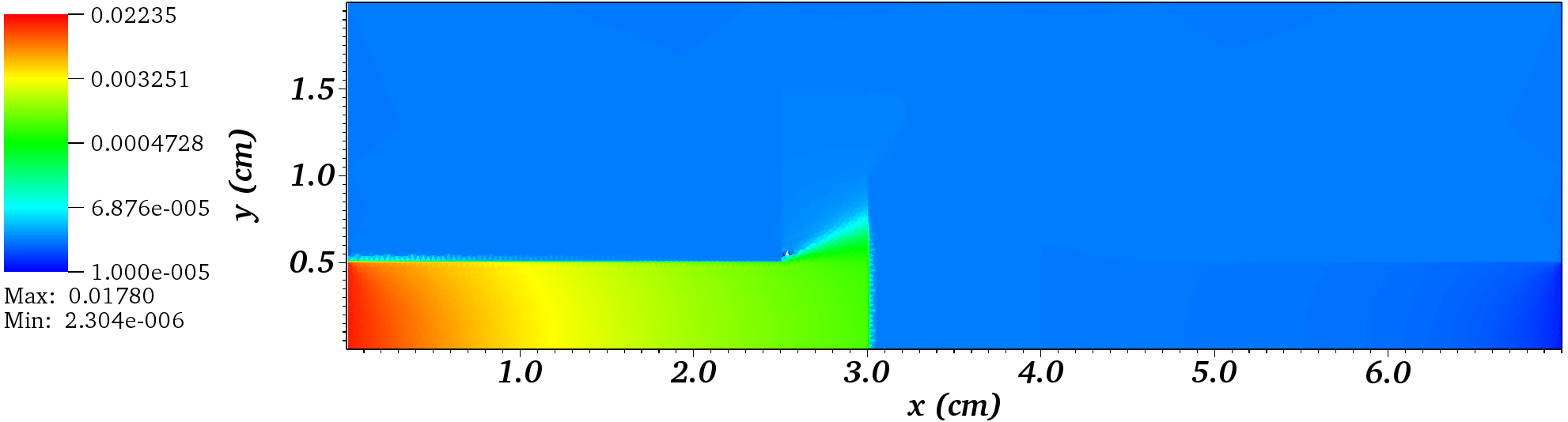}\\
	\caption{Scalar intensity $I_0^0$ (in GJ/cm$^2$/sh) at $t = 0.05$ sh for unfiltered P$_{1}$, P$_{3}$, P$_{5}$, and P$_{7}$ calculations (from top to bottom). The last plot is a uniformly filtered P$_{39}$ calculation for reference. The white regions show where $I_0^0$ is less than 10$^{-5}$ (i.e.~essentially negative with such a log scale). Only the piecewise constant component of the solution is shown.}
	\label{fig:2Dunfilt}
\end{figure}

\begin{figure}[H]
	\centering
	\includegraphics[width=0.99\textwidth]{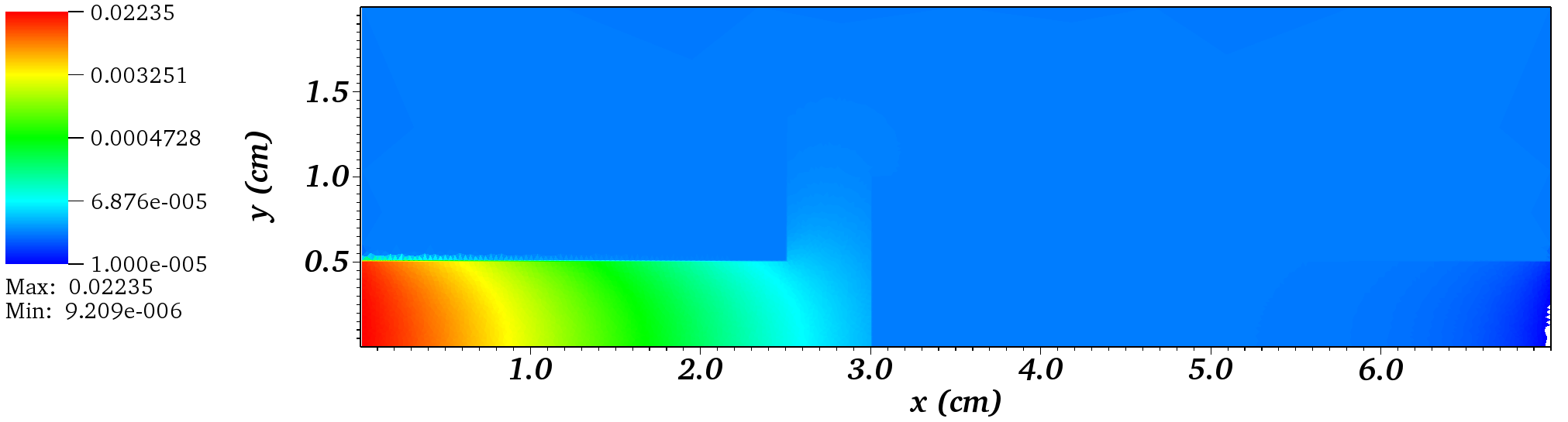}\\
	\includegraphics[width=0.99\textwidth]{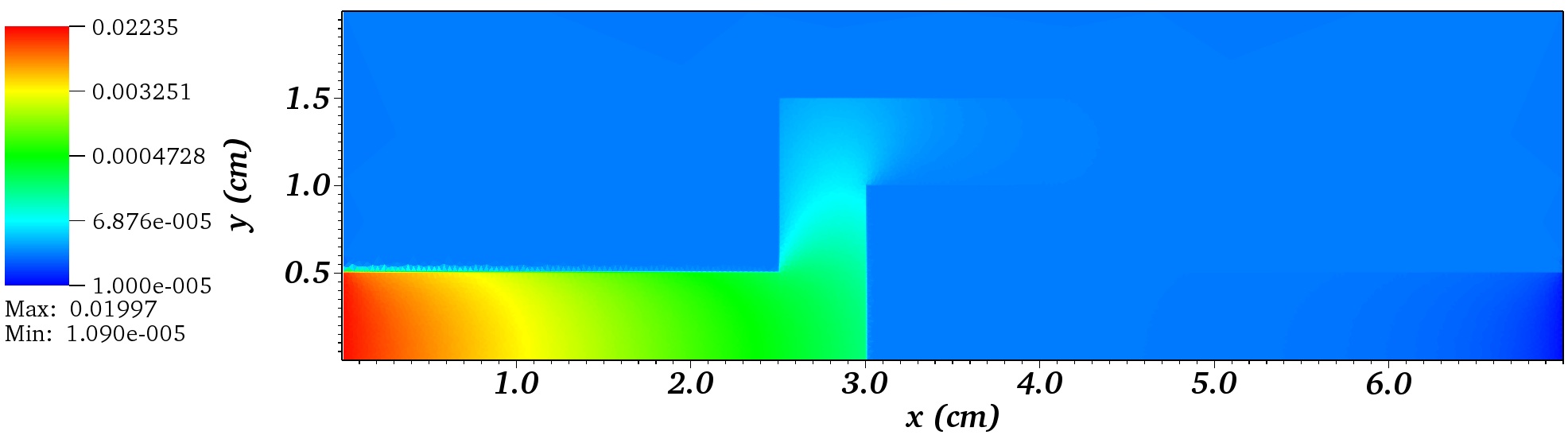}\\
	\includegraphics[width=0.99\textwidth]{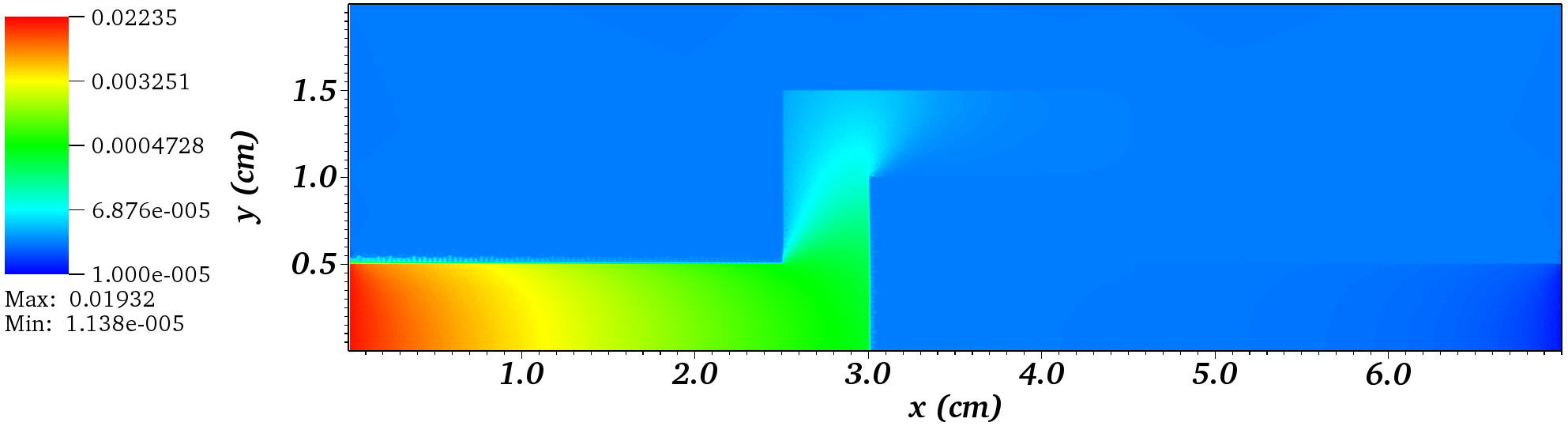}\\
	\includegraphics[width=0.99\textwidth]{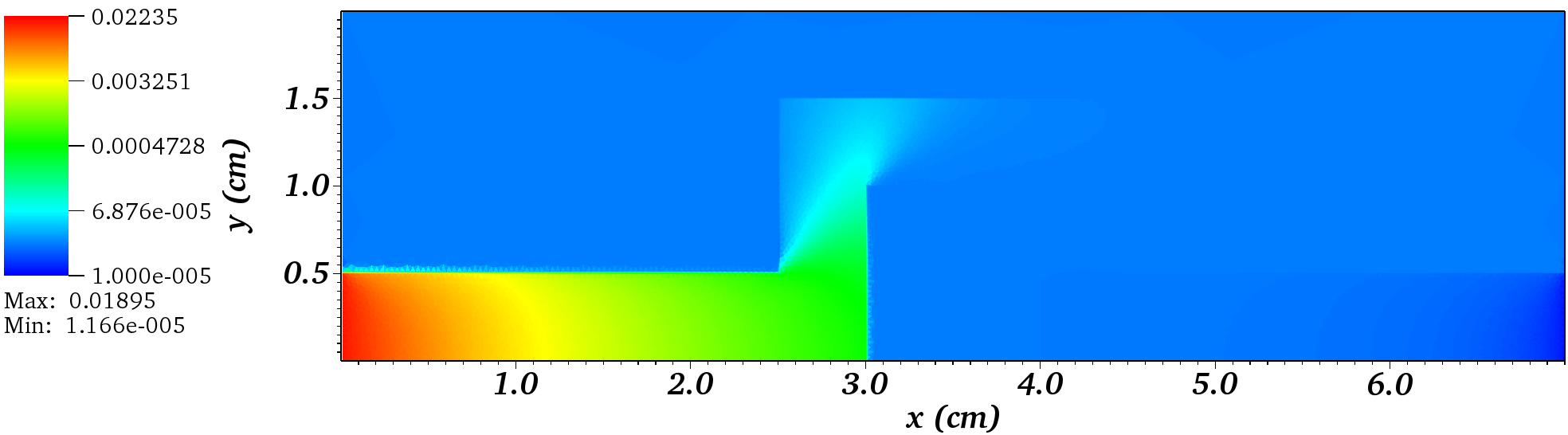}\\
	\includegraphics[width=0.99\textwidth]{phi_P39unif.png}\\
	\caption{Scalar intensity $I_0^0$ (in GJ/cm$^2$/sh) at $t = 0.05$ sh for uniformly filtered P$_{1}$, P$_{3}$, P$_{5}$, and P$_{7}$ calculations  (from top to bottom). The last plot is a uniformly filtered P$_{39}$ calculation for reference.  Only the piecewise constant component of the solution is shown.}
	\label{fig:2Dunif}
\end{figure}

\begin{figure}[H]
	\centering
	\includegraphics[width=0.99\textwidth]{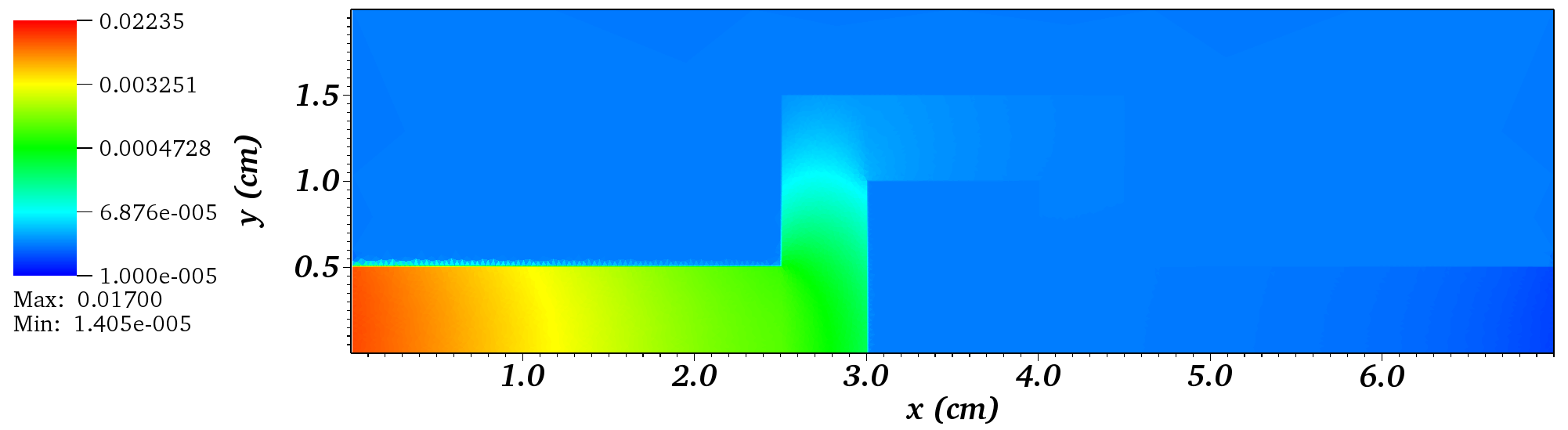}\\
	\includegraphics[width=0.99\textwidth]{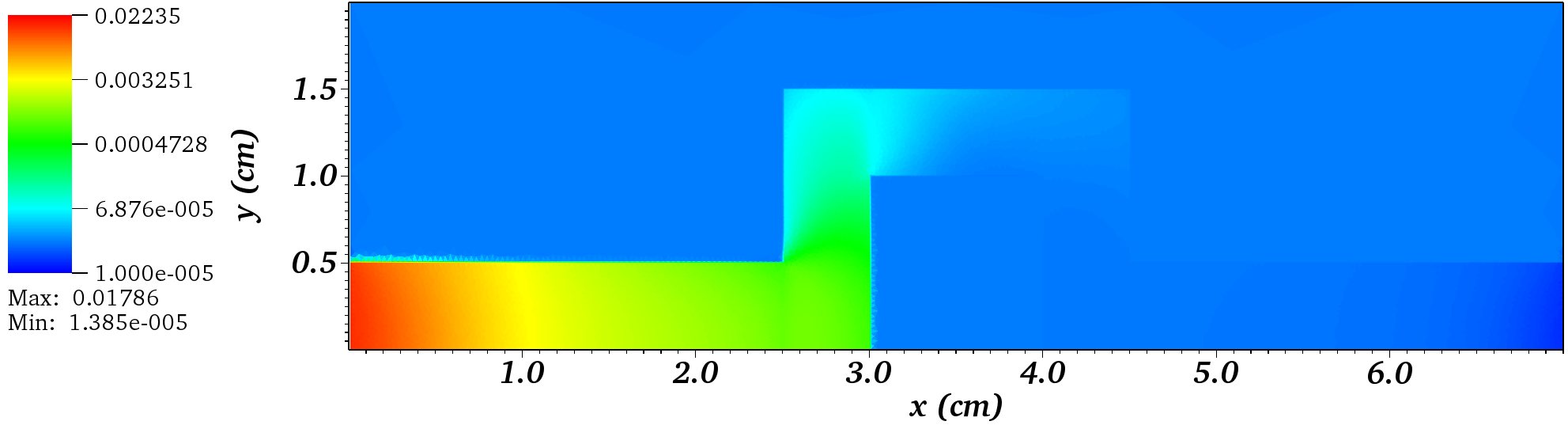}\\
	\includegraphics[width=0.99\textwidth]{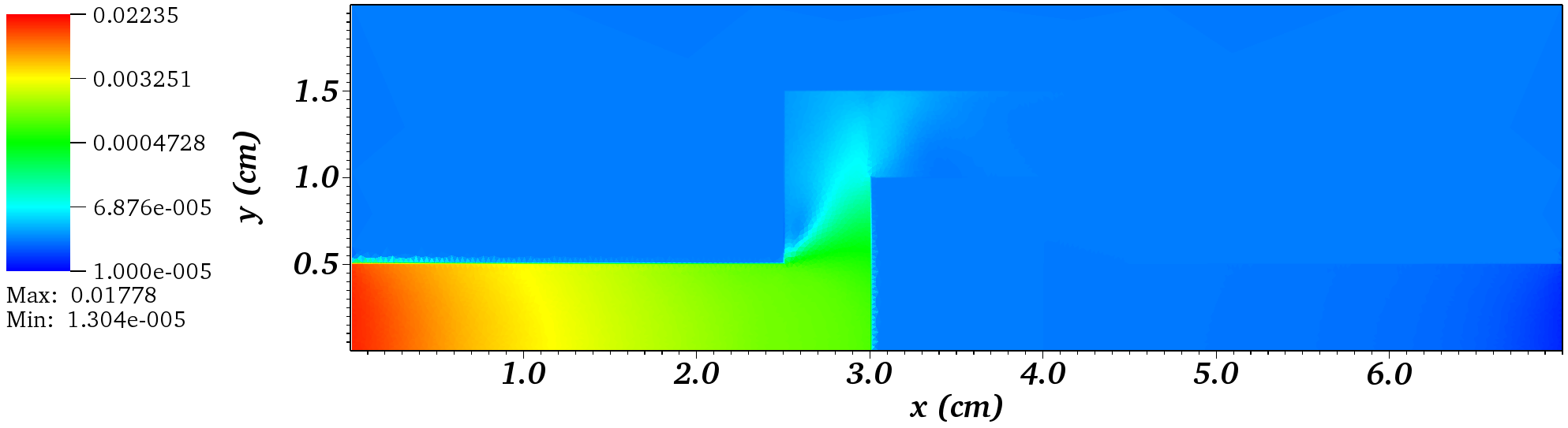}\\
	\includegraphics[width=0.99\textwidth]{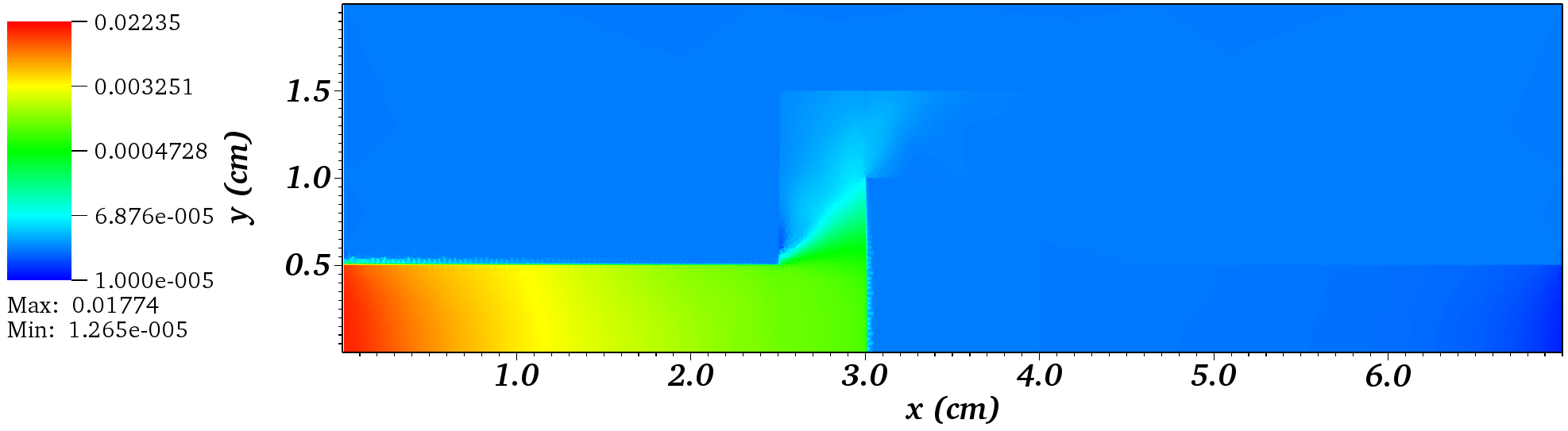}\\
	\includegraphics[width=0.99\textwidth]{phi_P39unif.png}\\
	\caption{Scalar intensity $I_0^0$ (in GJ/cm$^2$/sh) at $t = 0.05$ sh for locally filtered P$_{1}$, P$_{3}$, P$_{5}$, and P$_{7}$ calculations (from top to bottom). The last plot is a uniformly filtered P$_{39}$ calculation for reference.  Only the piecewise constant component of the solution is shown.}
	\label{fig:2DlocFilt}
\end{figure}

In Figs.\,\ref{fig:2Dunfilt}-\ref{fig:2DlocFilt}, we plot heat maps of the scalar intensity $I_0^0$ for the unfiltered, uniformly filtered, and locally filtered spherical harmonic calculations, respectively, at time $t=0.05$ sh.  It is around this time that the value of $I_0^0$ in the unfiltered solution reaches its minimum.  Each figure contains solutions for $N=1$, 3, 5 and $7$. The filtered P$_{39}$ solution with uniform filtering is included for reference.

Fig.\,\ref{fig:2Dunfilt} shows the defects of the P$_{N}$ closures. P$_1$ allows energy to flow through the thin region around the bend in the pipe. Meanwhile, the P$_{3}$, P$_{5}$ and P$_{7}$ calculations have regions -- the edge of shadows -- where the scalar intensity becomes negative. 
If a low enough initial temperature is chosen, the temperature will actually become negative, then yielding nonsensical results. Fig.\,\ref{fig:2Dunif} shows that uniform filtering efficiently removes regions of negativity, but also over-damps the scalar intensity profile for low values of $N$.


\subsubsection{Lineouts}
\label{SecLineouts}

\begin{figure}[H]
	\begin{subfigure}[b]{0.47\textwidth}
		\includegraphics[width=\textwidth]{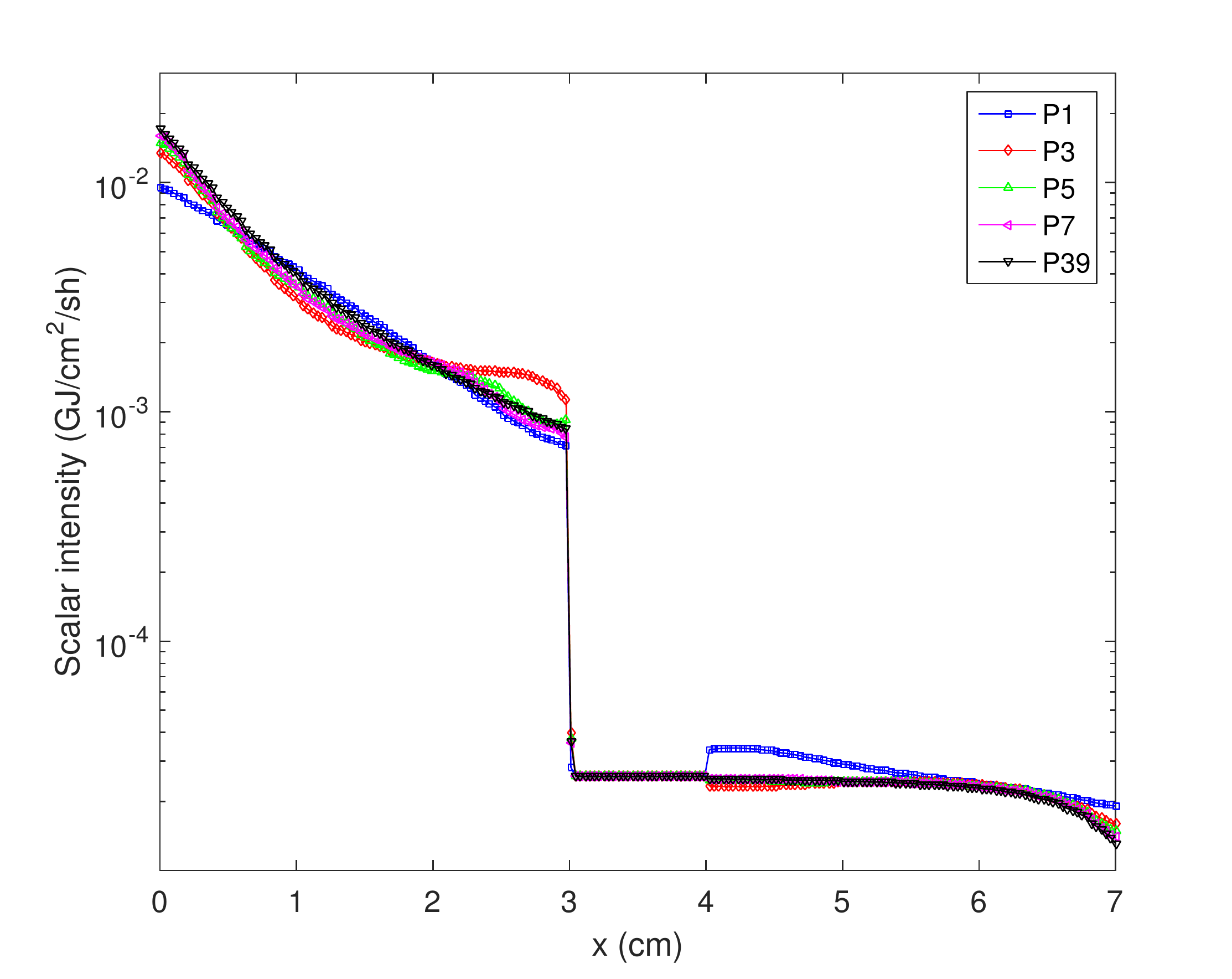}\\
		\caption{Unfiltered P$_N$.}
		\label{fig:y=0_b}
	\end{subfigure}%
	\quad 
	\begin{subfigure}[b]{0.47\textwidth}
		\includegraphics[width=\textwidth]{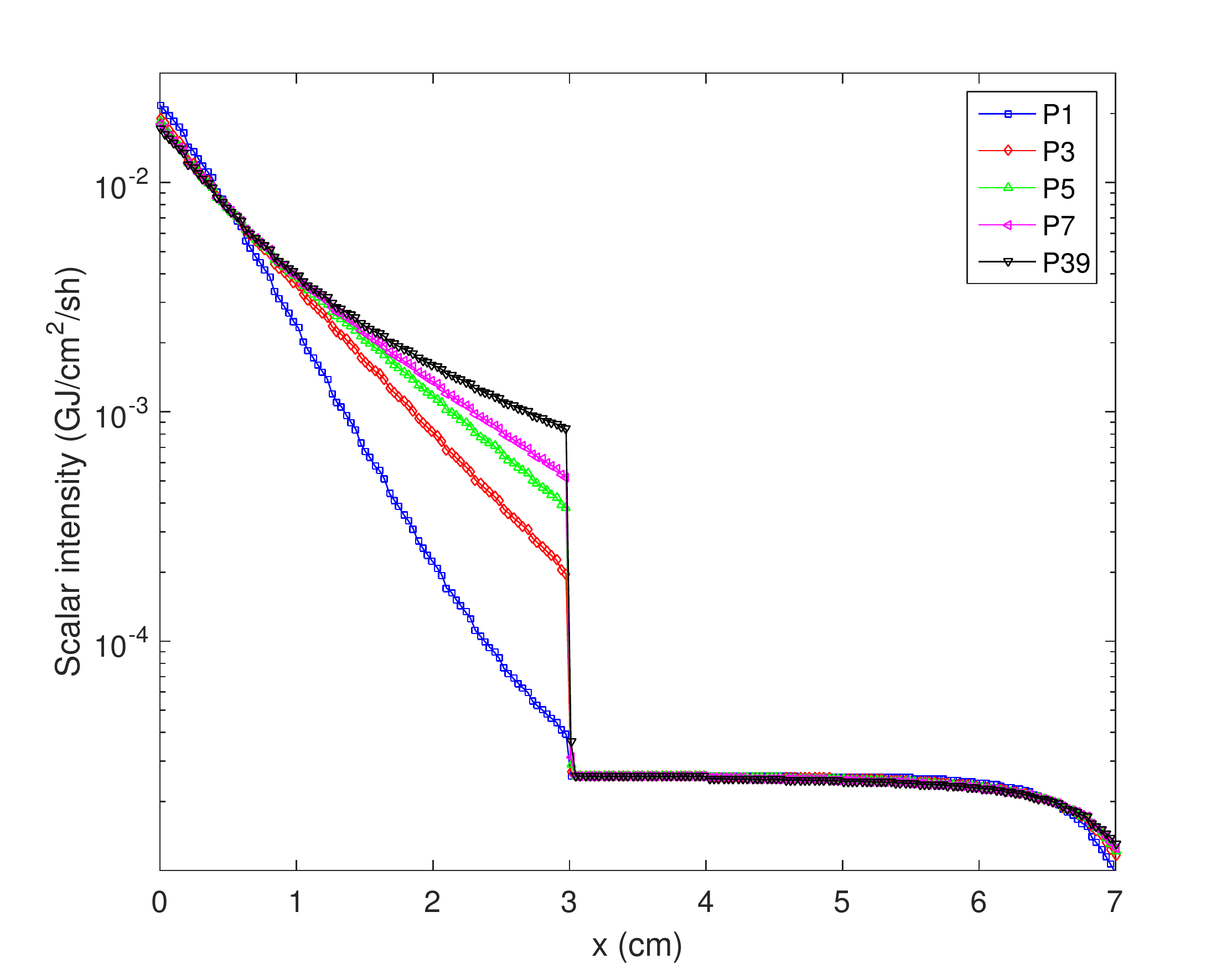}\\
		\caption{Uniformly Filtered P$_N$.}
		\label{fig:y=0_c}
	\end{subfigure}
	\vspace{0cm}\\
	\begin{subfigure}[b]{0.47\textwidth}
		\includegraphics[width=\textwidth]{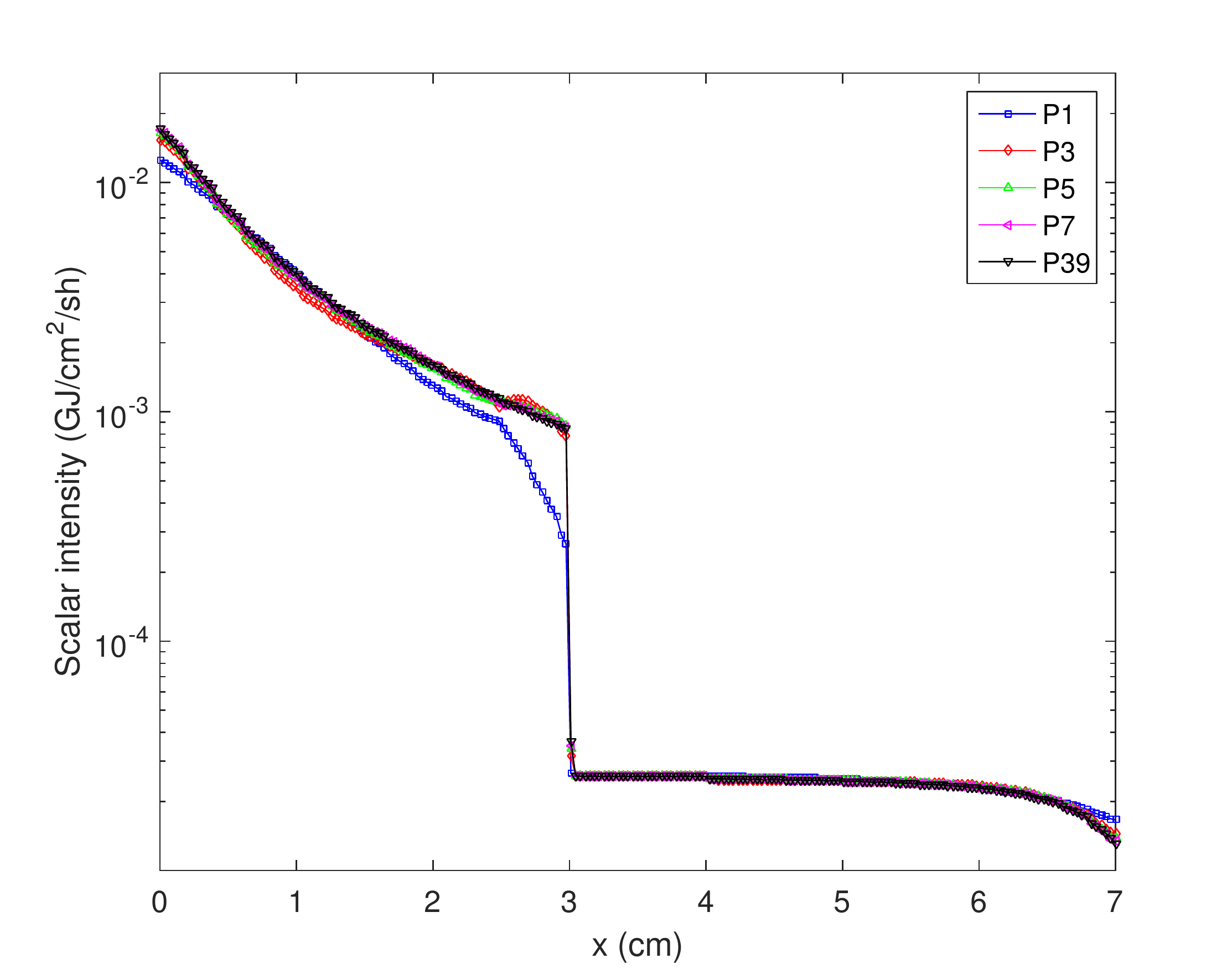}\\
		\caption{Locally Filtered P$_N$.}
		\label{fig:y=0_d}
	\end{subfigure}%
	\begin{subfigure}[b]{0.47\textwidth}
		\hspace*{0.5cm}
		\includegraphics[width=\textwidth]{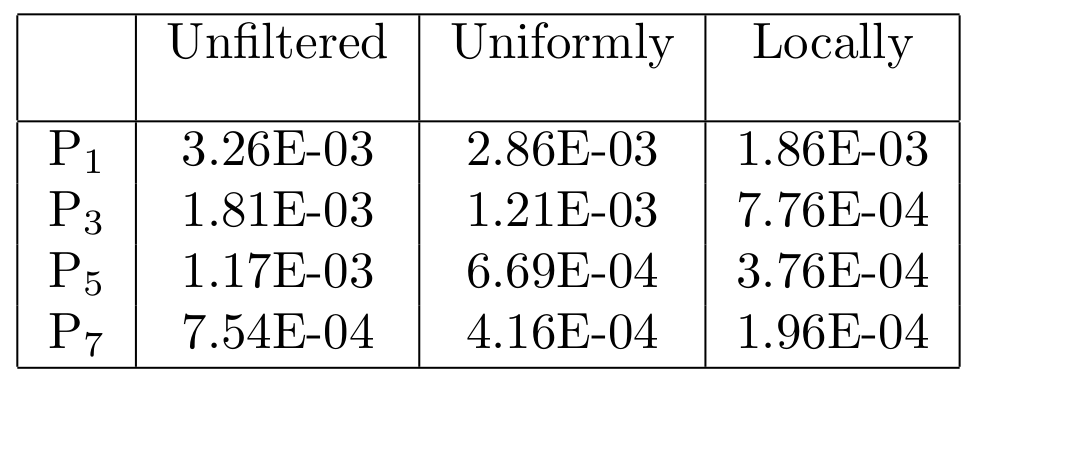}\\
		\caption{L2-error in GJ-cm$^{-3/2}$-sh$^{-1}$.}
		\label{fig:y=0_e}
	\end{subfigure}%
	\captionsetup{margin=0.5cm}
	\caption{Scalar intensity profile along the straight line $y = 0$ at $t=0.05$ sh (refer to Fig.\,\ref{fig2} to see where the straight line is with respect to the geometry). The stair-casing is an artifact of the visualization software, which plots piece-wise constants. 
	} 
	\label{fig:y=0}
\end{figure}

\begin{figure}[ht!]
	\begin{subfigure}[b]{0.47\textwidth}
		\includegraphics[width=\textwidth]{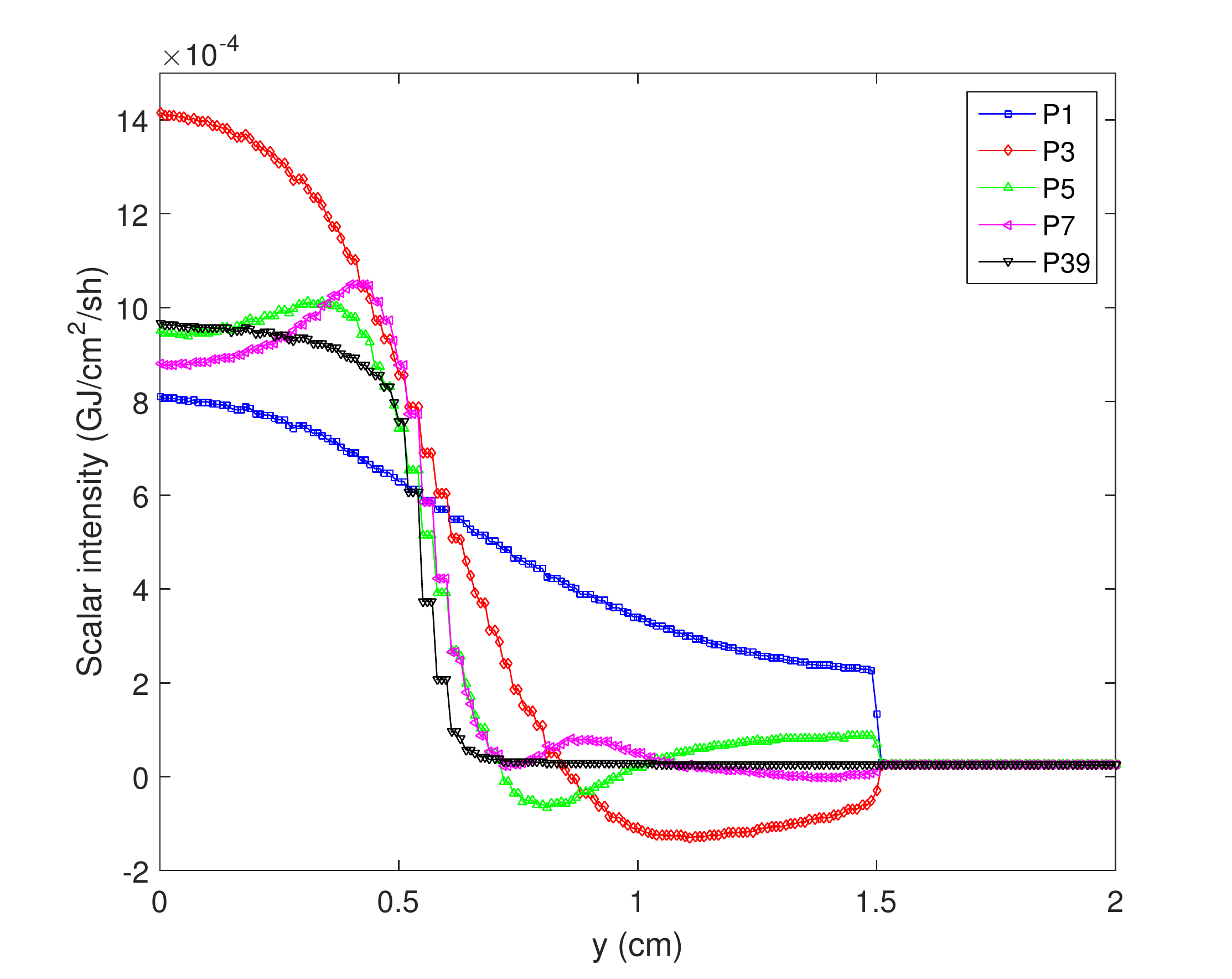}\\
		\caption{Unfiltered P$_N$.}
		\label{fig:x=2.75_b}
	\end{subfigure}%
	\quad 
	\begin{subfigure}[b]{0.47\textwidth}
		\includegraphics[width=\textwidth]{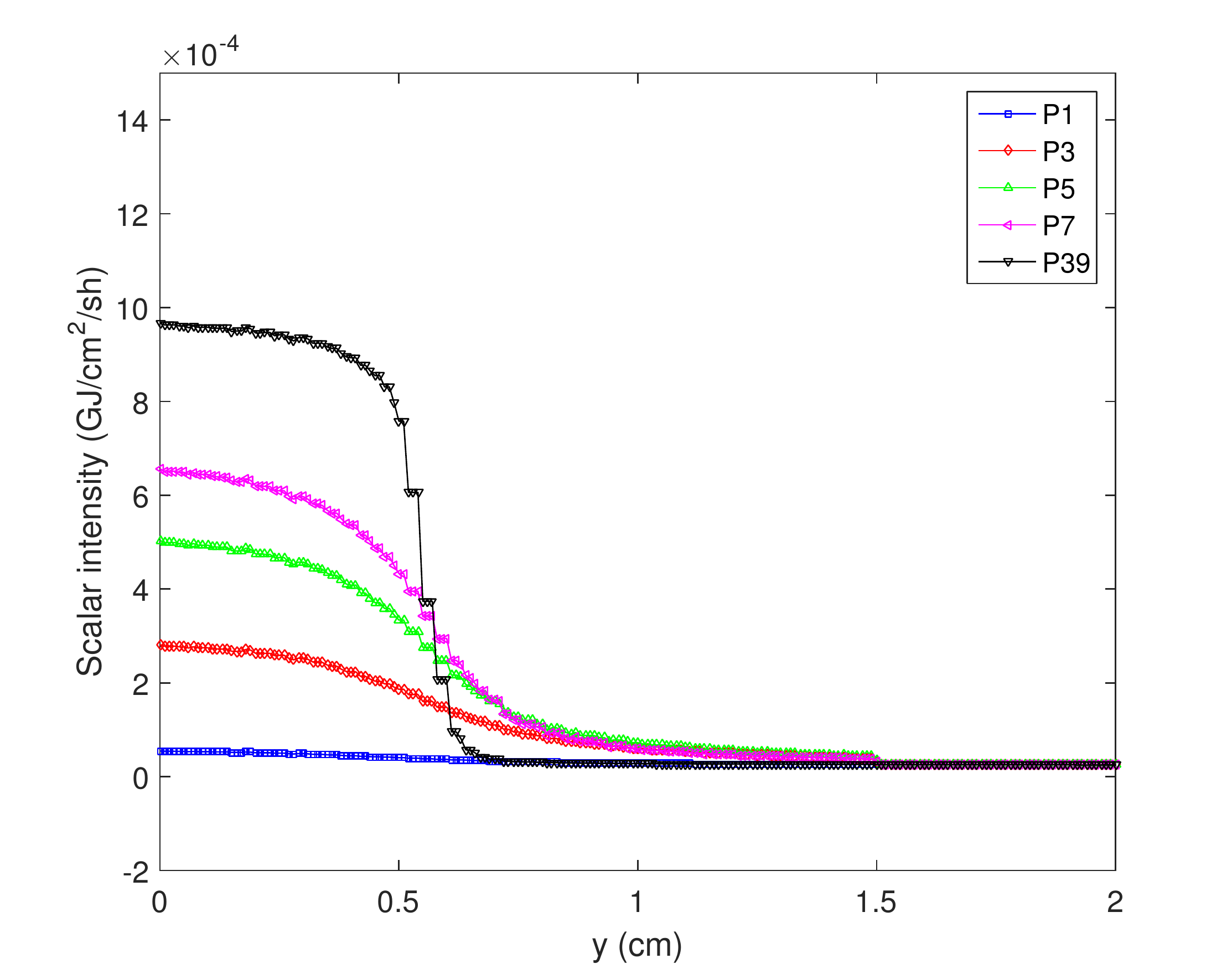}\\
		\caption{Uniformly Filtered P$_N$.}
		\label{fig:x=2.75_c}
	\end{subfigure}
	\vspace{0cm}\\
	\begin{subfigure}[b]{0.47\textwidth}
		\includegraphics[width=\textwidth]{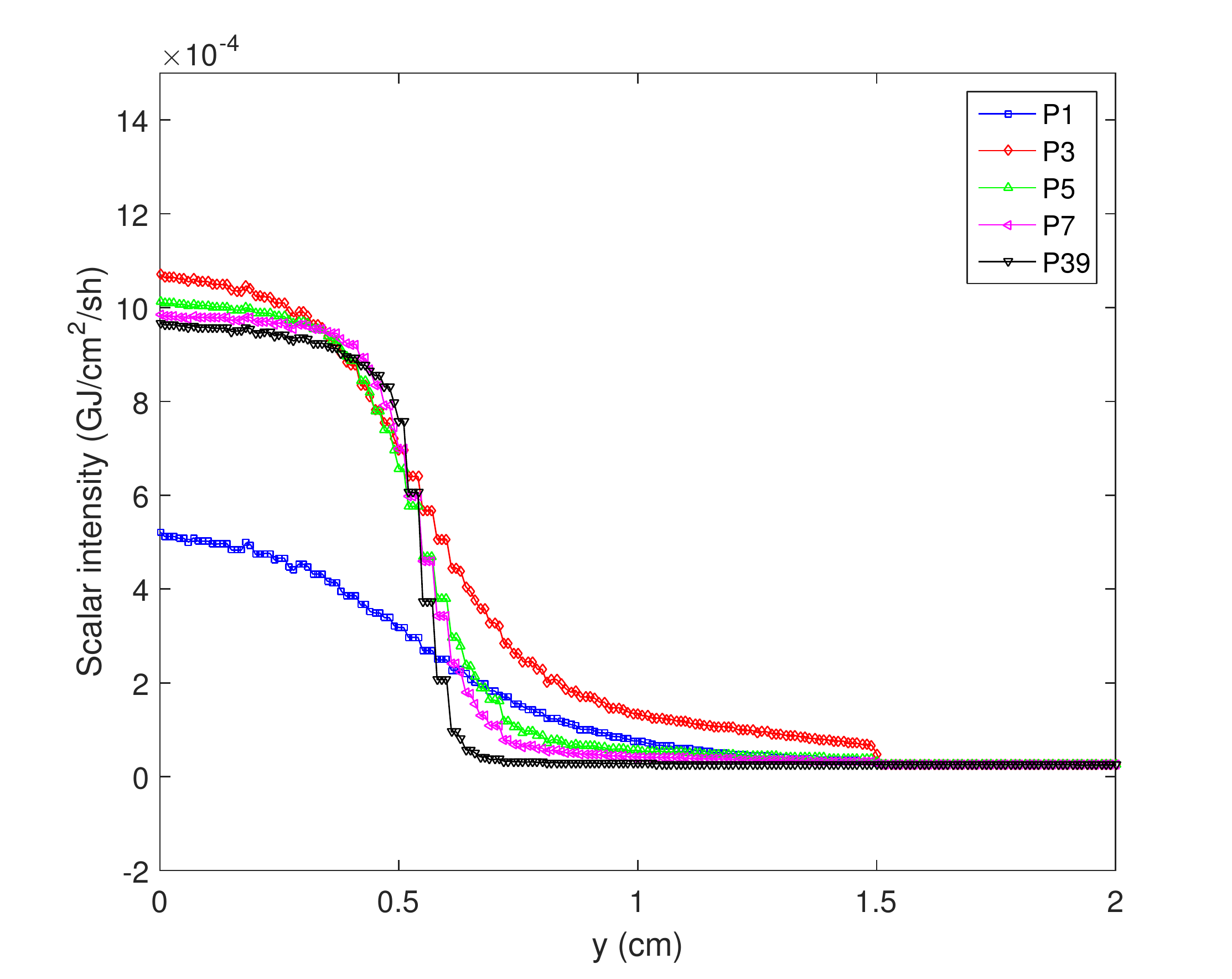}\\
		\caption{Locally Filtered P$_N$.}
		\label{fig:x=2.75_d}
	\end{subfigure}
	\begin{subfigure}[b]{0.47\textwidth}
		\hspace*{0.5cm}
		\includegraphics[width=\textwidth]{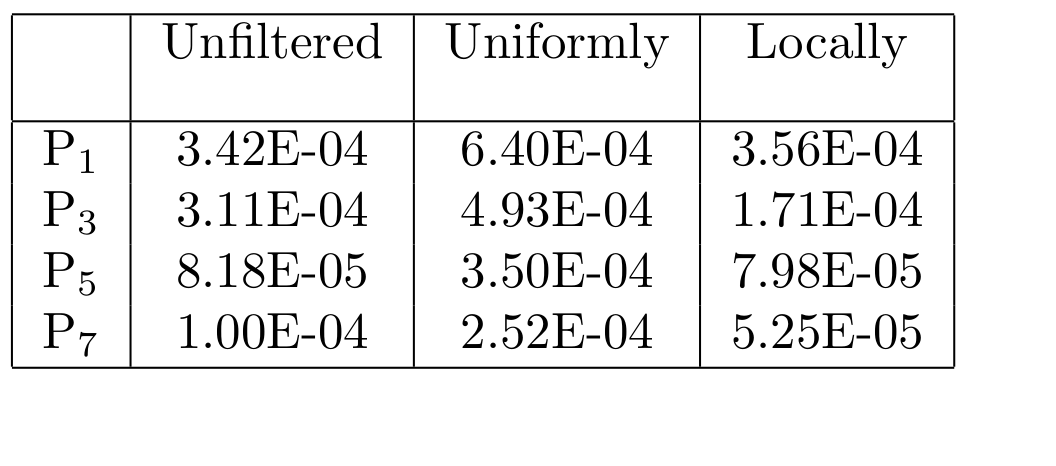}\\
		\caption{L2-error in GJ-cm$^{-3/2}$-sh$^{-1}$.}
		\label{fig:x=2.75_e}
	\end{subfigure}%
	\captionsetup{margin=0.5cm}
	\caption{Scalar intensity profile along the straight line $x = 2.75$ cm at $t=0.05$ sh (refer to Fig.\,\ref{fig2} to see where the straight line is with respect to the geometry). The stair-casing is an artifact of the visualization software, which plots piece-wise constants. 
	} 
	\label{fig:x=2.75}
\end{figure}

\addition{In this section and the following, we provide L2-error tables to quantify the filter performances. It is generally defined as $(\int_{u_{\min}}^{u_{\max}}(I_0^0-I_{\text{ref}})^2\,\mathrm{d}u )^{1/2}$, the reference being the P$_{39}$ curve. For the lineouts, $u$ represents the corresponding spatial variable ($u=x$ for Fig.~\ref{fig:y=0}, $u=y$ for Fig.~\ref{fig:x=2.75}). For the time histories, it represents the time $t$.}

Figs.\,\ref{fig:y=0} and \ref{fig:x=2.75} show lineouts of the scalar intensity profile at time $t=0.05$ sh along the lines $y=0$ cm and $x=2.75$ cm, respectively.
Except for P$_1$, all of the unfiltered P$_N$ solutions (Fig.\,\ref{fig:y=0_b}) along $y=0$  are very similar and agree with the reference solution to within $12\%$.
In the uniformly filtered case (Fig.\,\ref{fig:y=0_c}), over damping has slowed the effective flow of radiation down the pipe, causing solutions to be much less accurate. Meanwhile, the locally filtered results (Fig.\,\ref{fig:y=0_d}) are slightly better than the unfiltered ones.

Along the line $x = 2.75$ cm, nonphysical oscillations cause the scalar intensity profile for the unfiltered equations (Fig.\,\ref{fig:x=2.75_b}) to reach negative values. The filter helps significantly in this region, with the local filter (Fig.\,\ref{fig:x=2.75_d}) again outperforming the uniform one, especially for small values of $N$. Even so, the filtered solutions do over-predict the scalar intensity compared to the P$_{39}$ solution after the first elbow.

\subsubsection{Time Histories}
\label{SecTimeplots}
\begin{figure}[H]
	\begin{subfigure}[b]{0.47\textwidth}
		\includegraphics[width=\textwidth]{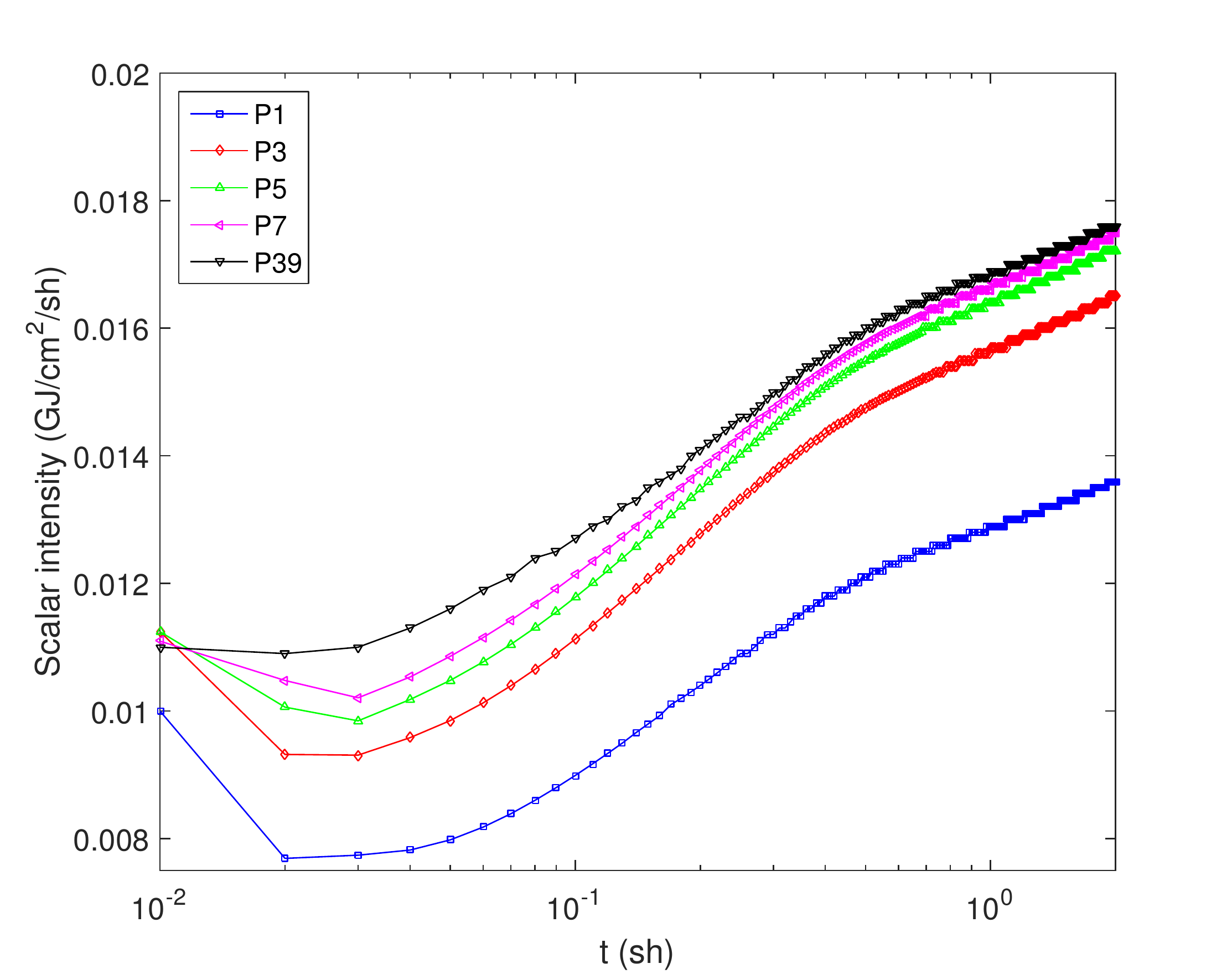}\\
		\caption{Unfiltered P$_N$.}
		\label{fig:T1_b}
	\end{subfigure}%
	\quad 
	\begin{subfigure}[b]{0.47\textwidth}
		\includegraphics[width=\textwidth]{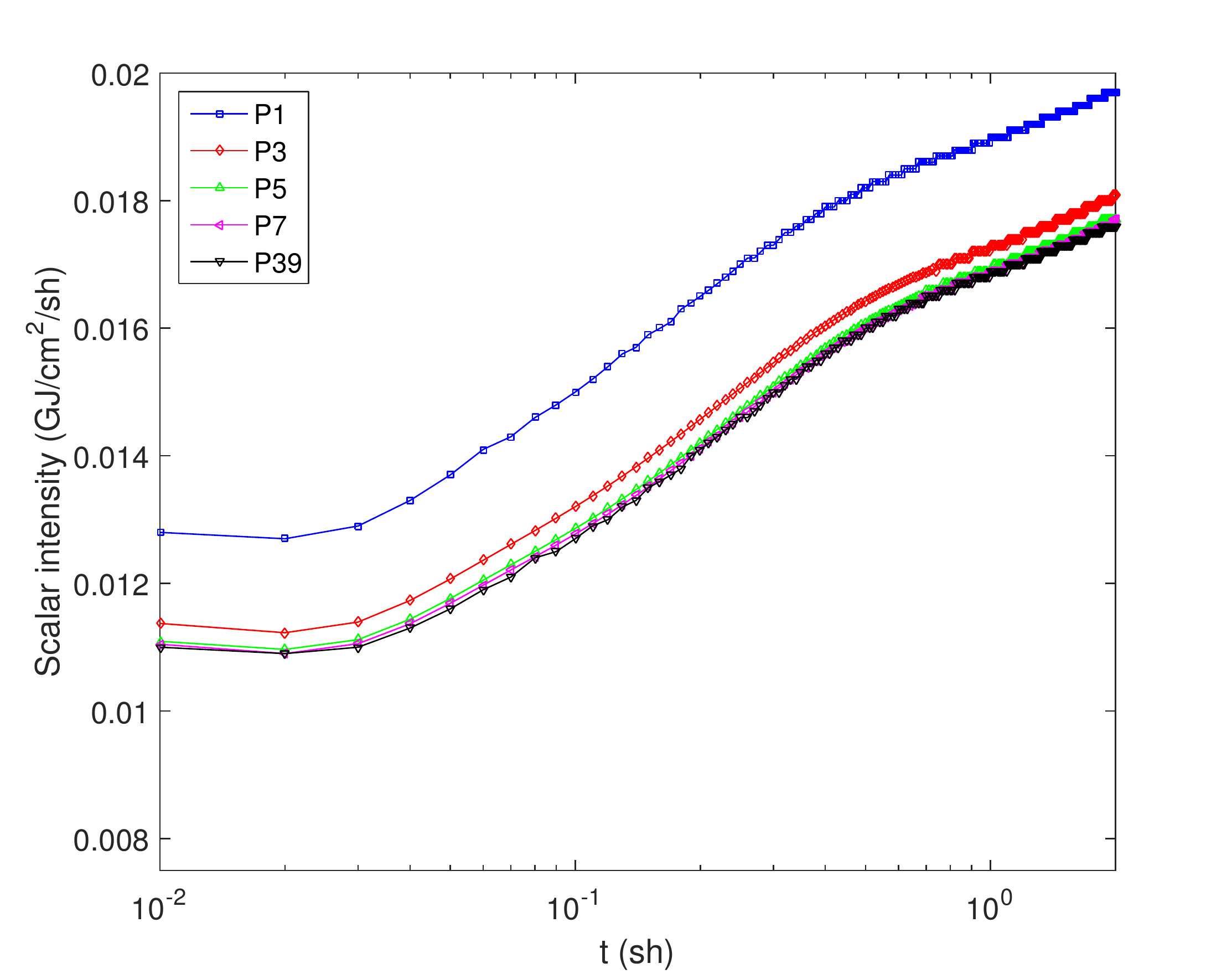}\\
		\caption{Uniformly Filtered P$_N$.}
		\label{fig:T1_c}
	\end{subfigure}
	\vspace{0cm}\\
	\begin{subfigure}[b]{0.47\textwidth}
		\includegraphics[width=\textwidth]{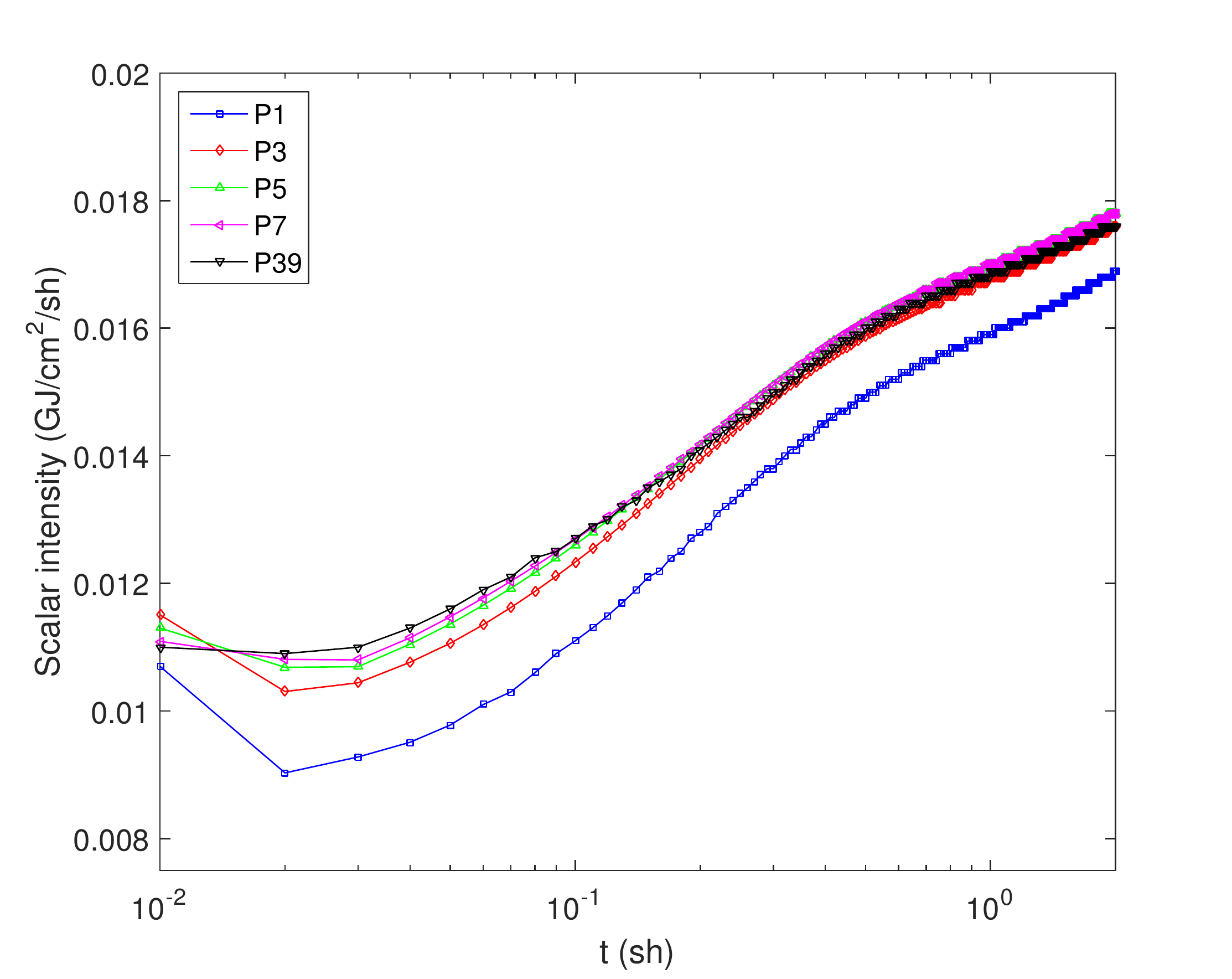}\\
		\caption{Locally Filtered P$_N$.}
		\label{fig:T1_d}
	\end{subfigure}
	\begin{subfigure}[b]{0.47\textwidth}
		\hspace*{0.5cm}
		\includegraphics[width=\textwidth]{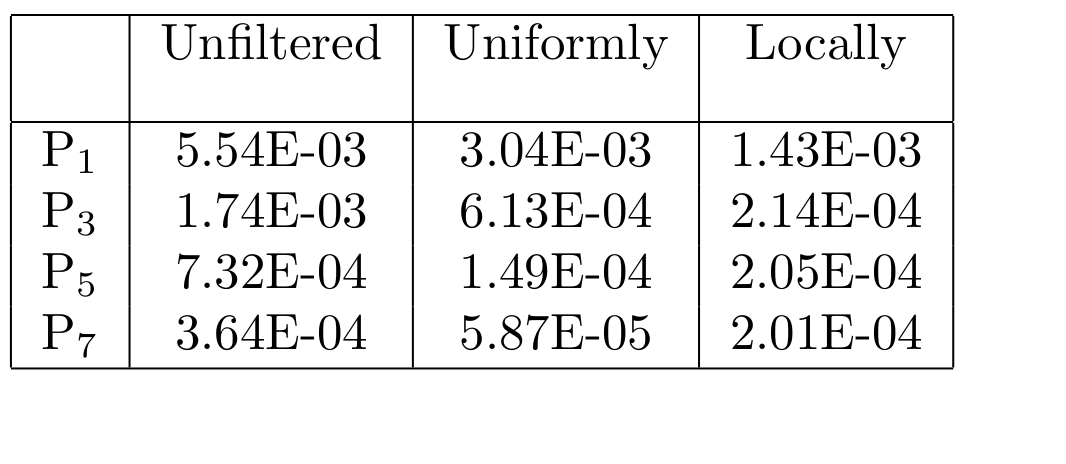}\\
		\caption{L2-error in GJ-cm$^{-2}$-sh$^{-1/2}$.}
		\label{fig:T1_e}
	\end{subfigure}%
	\captionsetup{margin=0.5cm}
	\caption{Scalar intensity profile at the point $(x_{1},y_{1})=$ (0.25\,cm,\,0). Refer to Fig.\,\ref{fig2} to see where this point lies with respect to the geometry. 
	} 
	\label{fig:T1}
\end{figure}

As suggested in \cite{CrookedPipe}, we also monitor the evolution of $I^0_0$ as a function of time at 3 different points in space: $(x_{1},y_{1})=$ (0.25\,cm,\,0), $(x_{2},y_{2})=$ (2.75\,cm,\,0), and $(x_{3},y_{3})=$ (3.5\,cm,\,1.25\,cm).  These results are given in Figs.\,\ref{fig:T1}\,-\,\ref{fig:T3}.  

At $(x_{1},y_{1})$ (Fig.\,\ref{fig:T1}), all the filtering approaches give reasonable results.
The values of $I^0_0$ for uniform filtering in Fig.\,\ref{fig:T1_c} are slightly higher than with the other two types because the radiation propagates more slowly and is therefore more concentrated at the entrance of the pipe. For the same reason, the unfiltered calculations tend to underestimate the temperature at that point for small values of $N$.

\begin{figure}
	\begin{subfigure}[b]{0.47\textwidth}
		\includegraphics[width=\textwidth]{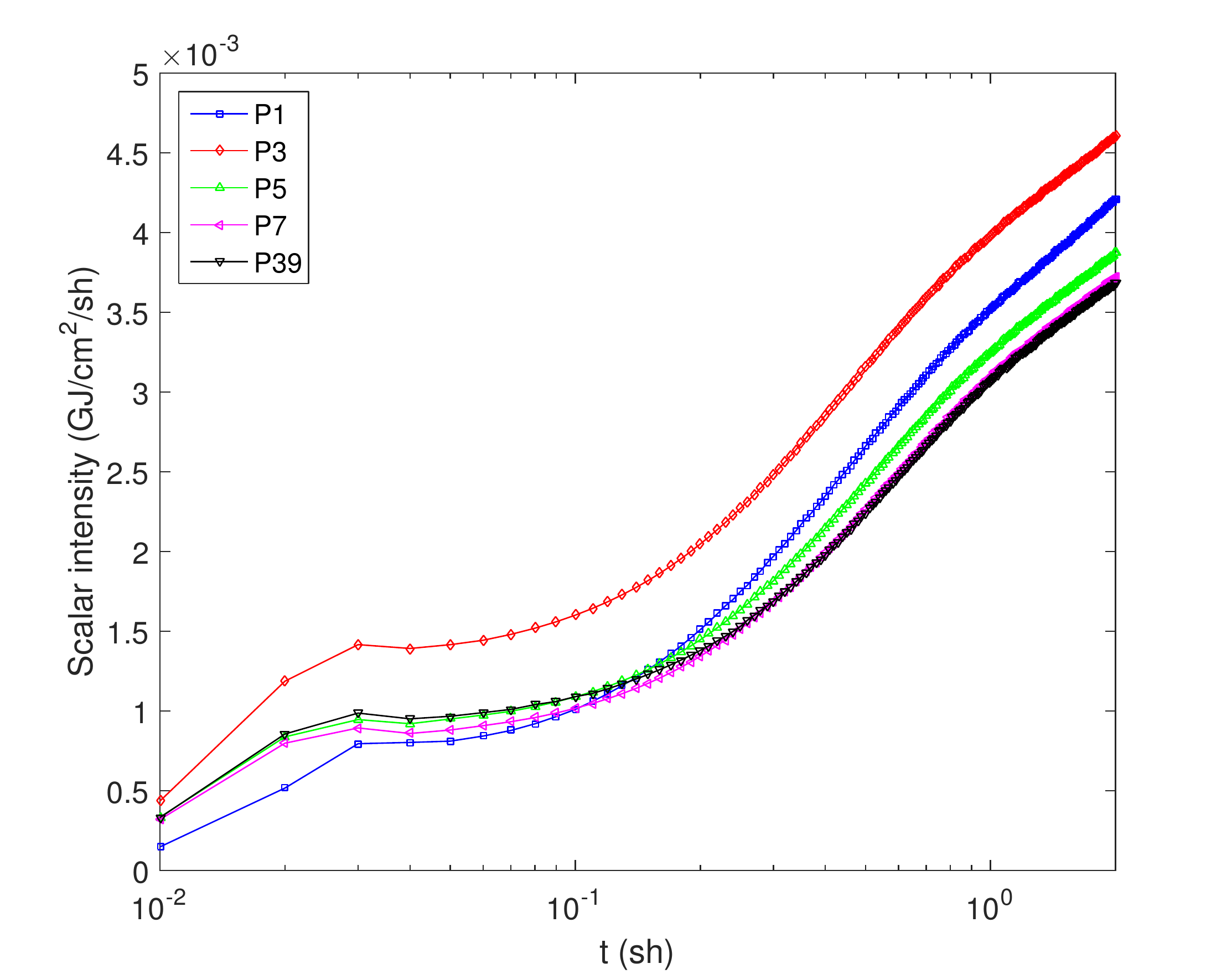}\\
		\caption{Unfiltered P$_N$.}
		\label{fig:T2_b}
	\end{subfigure}%
	\quad 
	\begin{subfigure}[b]{0.47\textwidth}
		\includegraphics[width=\textwidth]{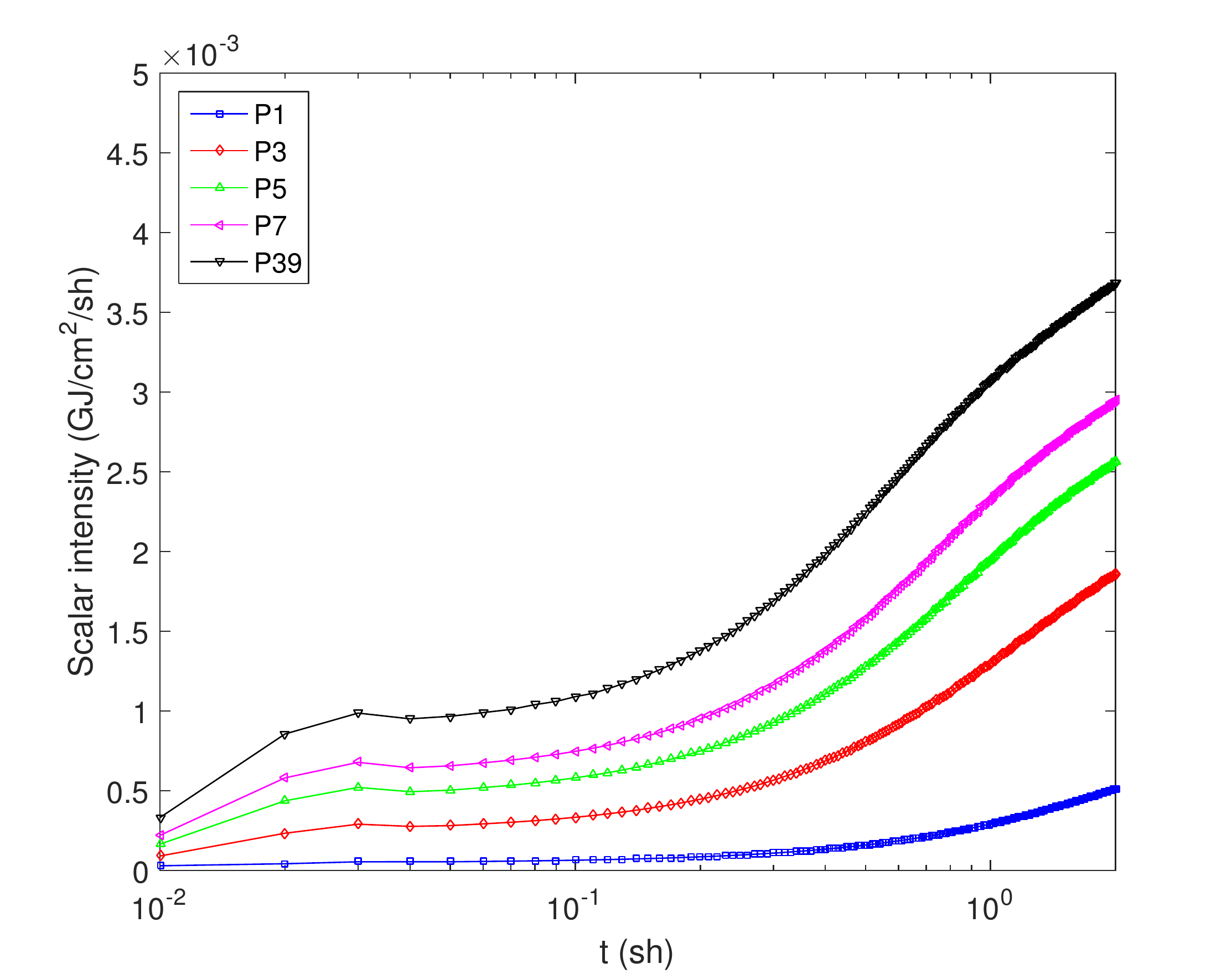}\\
		\caption{Uniformly Filtered P$_N$.}
		\label{fig:T2_c}
	\end{subfigure}
	\vspace{0cm}\\
	\begin{subfigure}[b]{0.47\textwidth}
		\includegraphics[width=\textwidth]{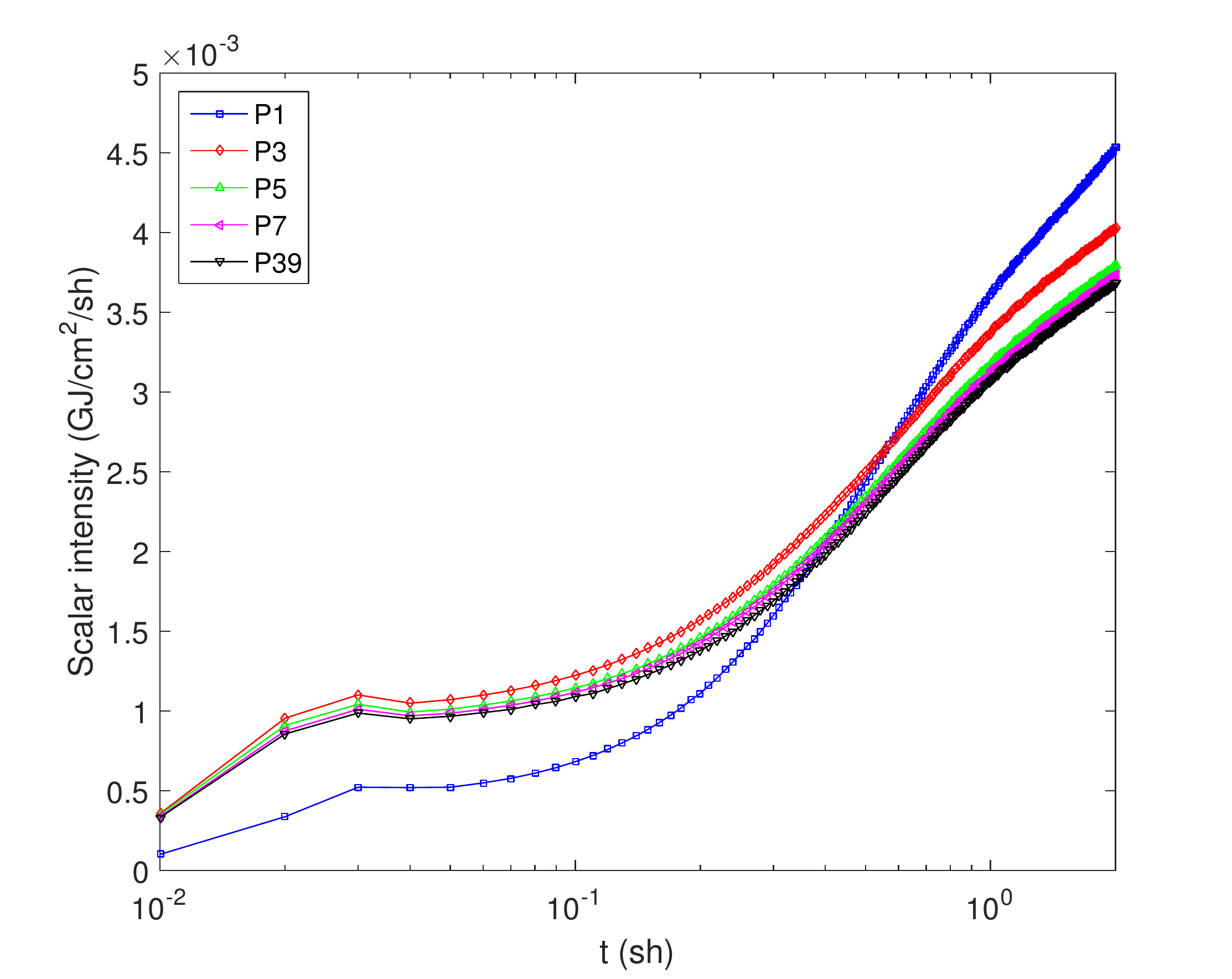}\\
		\caption{Locally Filtered P$_N$.}
		\label{fig:T2_d}
	\end{subfigure}
	\begin{subfigure}[b]{0.47\textwidth}
		\hspace*{0.5cm}
		\includegraphics[width=\textwidth]{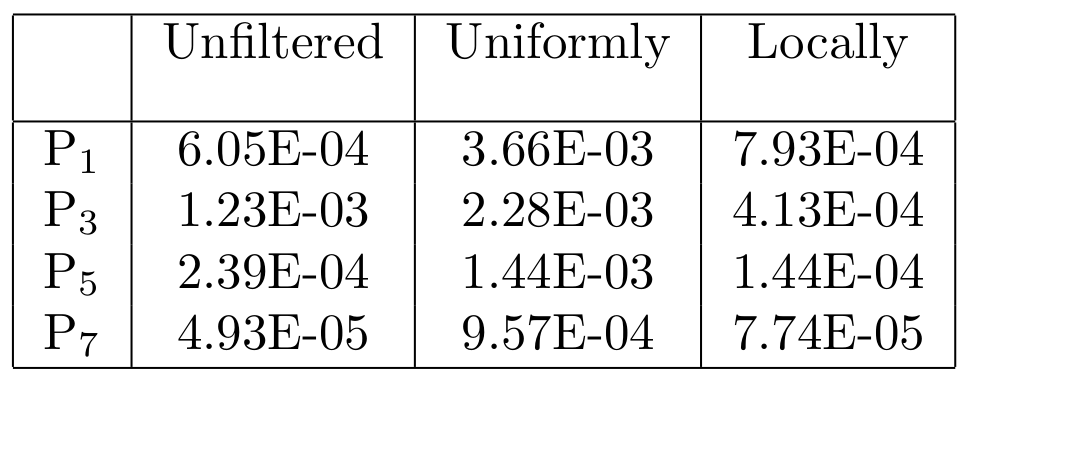}\\
		\caption{L2-error in GJ-cm$^{-2}$-sh$^{-1/2}$.}
		\label{fig:T2_e}
	\end{subfigure}%
	\captionsetup{margin=0.5cm}
	\caption{Scalar intensity profile at the point $(x_{2},y_{2})=$ (2.75\,cm,\,0). Refer to Fig.\,\ref{fig2} to see where this point lies with respect to the geometry.
	}
	\label{fig:T2}
\end{figure}

\begin{figure}
	\begin{subfigure}[b]{0.47\textwidth}
		\includegraphics[width=\textwidth]{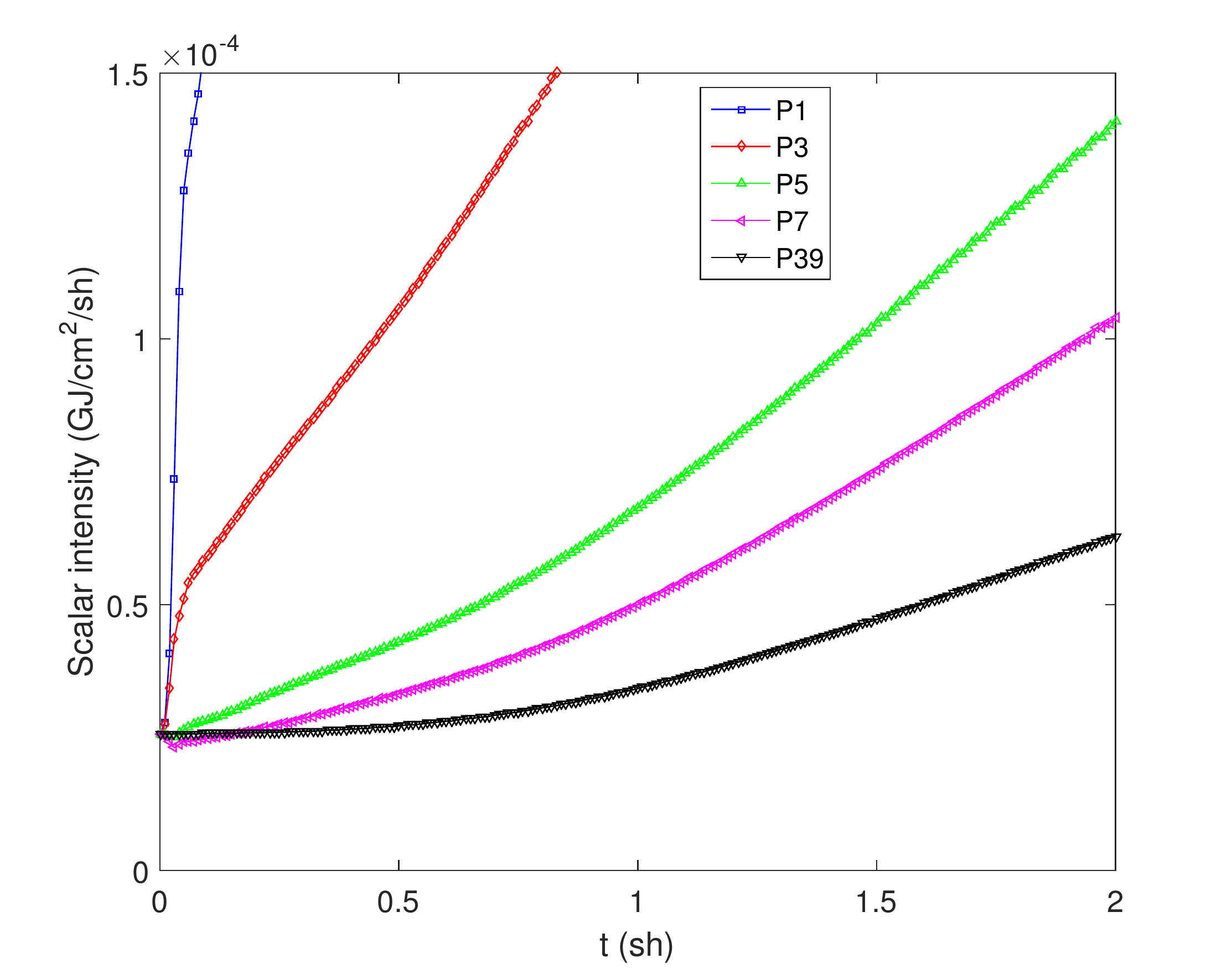}\\
		\caption{Unfiltered P$_N$.}
		\label{fig:T3_b}
	\end{subfigure}%
	\quad 
	\begin{subfigure}[b]{0.47\textwidth}
		\includegraphics[width=\textwidth]{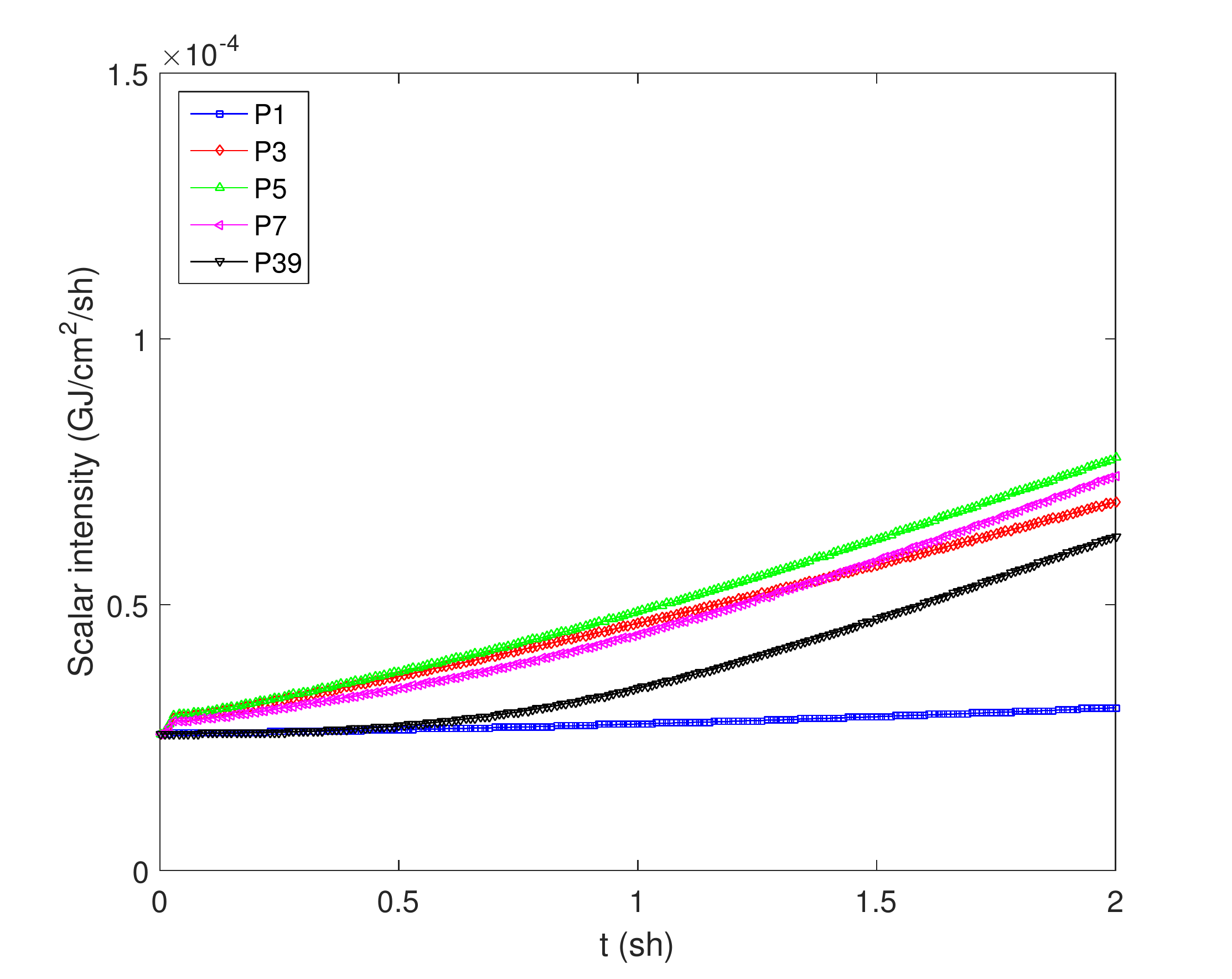}\\
		\caption{Uniformly Filtered P$_N$.}
		\label{fig:T3_c}
	\end{subfigure}
	\vspace{0cm}\\
	\begin{subfigure}[b]{0.47\textwidth}
		\includegraphics[width=\textwidth]{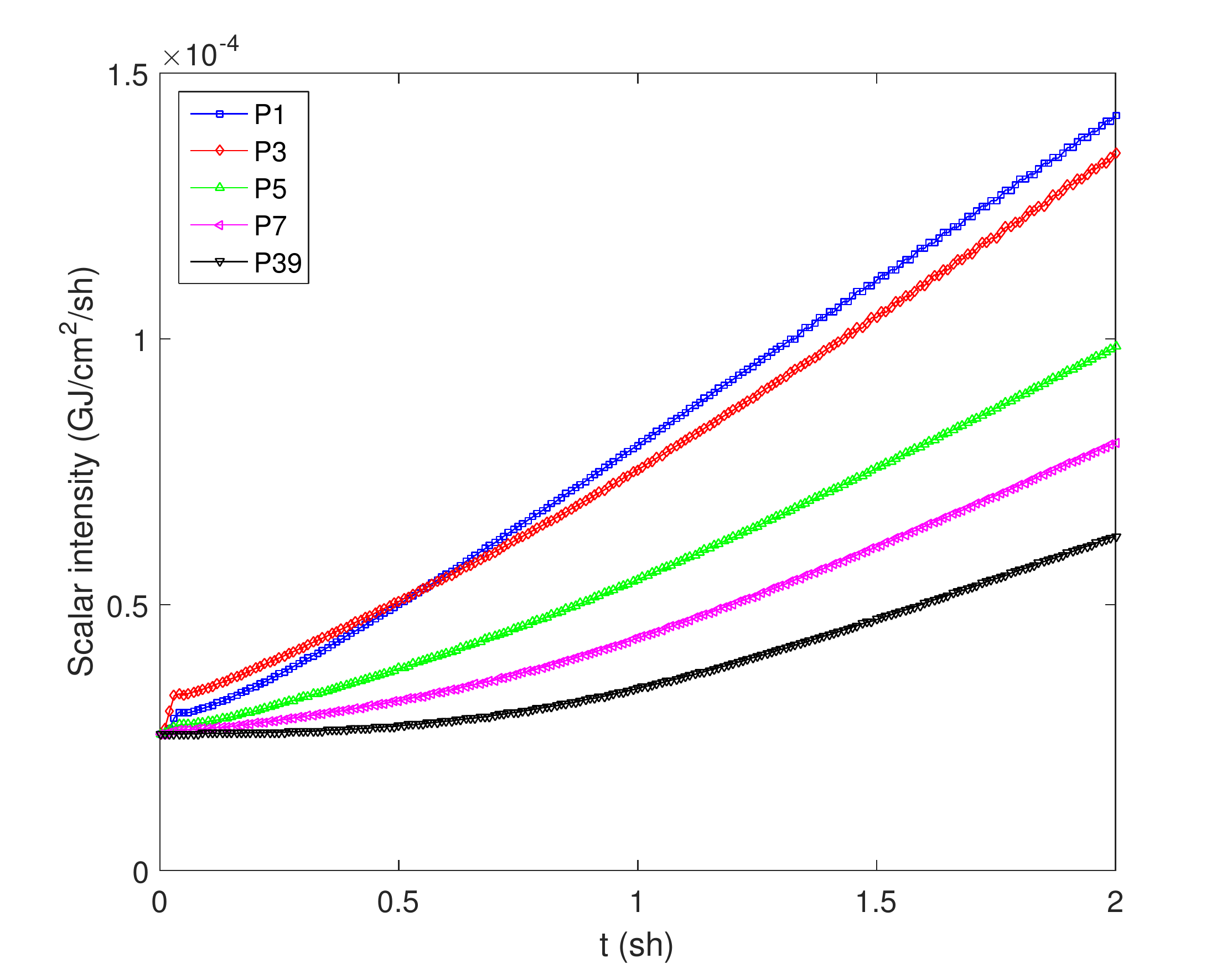}\\
		\caption{Locally Filtered P$_N$.}
		\label{fig:T3_d}
	\end{subfigure}
	\begin{subfigure}[b]{0.47\textwidth}
		\hspace*{0.5cm}
		\includegraphics[width=\textwidth]{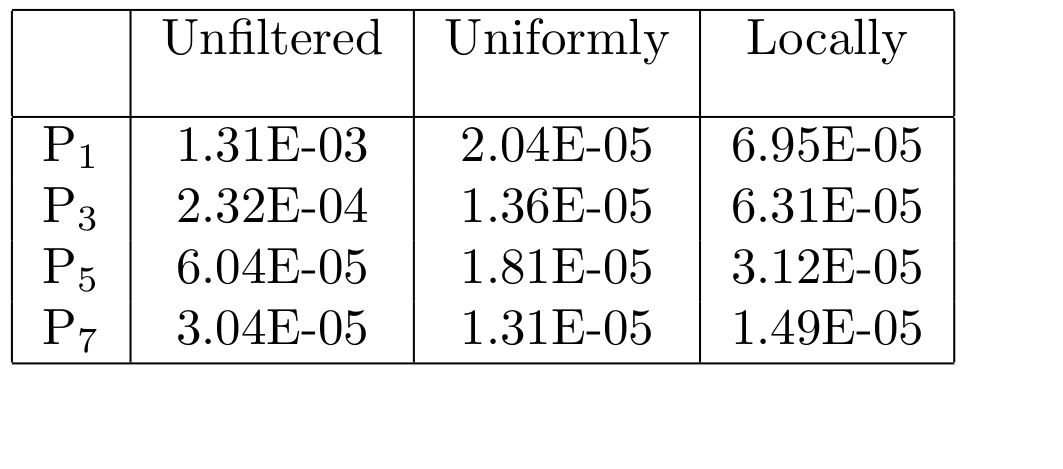}\\
		\caption{L2-error in GJ-cm$^{-2}$-sh$^{-1/2}$.}
		\label{fig:T3_e}
	\end{subfigure}%
	\captionsetup{margin=0.5cm}
	\caption{Scalar intensity profile at the point $(x_{3},y_{3})=$ (3.5\,cm,\,1.25\,cm). Refer to Fig.\,\ref{fig2} to see where this point lies with respect to the geometry.
	} 
	\label{fig:T3}
\end{figure}

At $(x_{2},y_{2})$, in Fig.\,\ref{fig:T2_b}, the unfiltered solutions are all reasonably close to the P$_{39}$ solution at early times (see Table\,\ref{fig:T2_e}), except for the P$_3$ solution, which is affected by the time history at this point.
Similar behavior for P$_5$ or P$_7$ can be observed at different points in space. The uniformly filtered solutions (Fig.\,\ref{fig:T2_c}) again suffer from over damping, while the locally filtered results (Fig.\,\ref{fig:T2_d}) agree well with the reference solution. Only P$_1$ does not capture the shape accurately.

At $(x_{3},y_{3})$ in Fig.\,\ref{fig:T3_b}, the unfiltered scalar intensities are too high.  The filtering improves this, with the uniform filter giving the best results for $N=1$ and $N=3$.  For $N=5$ and $N=7$, the local and uniform filters have similar errors.

\newpage
\section{Conclusions}
We have presented and implemented a fully-implicit, discontinuous Galerkin finite element method  for simulating filtered spherical harmonic (P$_{N}$) equations in the context of thermal radiative transfer and provided guidelines to determine filtering strategies for general problems. 
Interestingly, the conditioning of underlying linear systems improves for moderate values of the filter strength $\sigf$.  Indeed, it was observed that such values led to a significant reduction in the number of {GMRES} iterations needed to solve the Crooked Pipe benchmark problem.  We have also tested numerically the convergence properties of the filter and have found that the properties of the linear, pure transport problem carry over to the non-linear, thermal problem.    Roughly speaking, the filter order determines the convergence rate for smooth solutions, while for non-smooth problems, the filter has little impact.  Finally, we have performed detailed simulations of the  Crooked Pipe problem and used it as a test case to compare different filtering strategies.  We observe that filtering improves numerical solutions significantly, especially for small values of $N$. For the most part, it is a local filtering strategy that works best.

In the future, we wish to extend this work to problems with multiple energy groups.  In addition, we will apply the filter to second-order forms of the transport equation that are commonly used in the neutronics community.

\section*{Acknowledgments}
We are very thankful to Dr.\,Alex Long and Anthony Barbu for their help in comparing our FP$_N$ code to their IMC and discrete ordinate (S$_N$) codes, respectively.

\bibliographystyle{elsarticle-num}
\bibliography{paperDraftBib}

\end{document}